\documentclass[intlimits,twoside,a4paper]{article}

\usepackage[cp1251]{inputenc}

\usepackage[eqsecnum]{cmpj3}

\issue{2024}{27}{2}{23602}
\doinumber{10.5488/CMP.27.23602}

\title[Phase behaviour of ionic liquids in disordered porous media]%
{Vapour-liquid phase behaviour of primitive models of ionic liquids confined in disordered porous media%
}
\author[T. Hvozd, T. Patsahan, Yu. Kalyuzhnyi, O. Patsahan, M. Holovko]{T. Hvozd\orcid{0000-0002-0156-1753}\refaddr{label1}\thanks{Corresponding author: \email{tarashvozd@gmail.com, tarashvozd@icmp.lviv.ua}.}, 
        T. Patsahan\orcid{0000-0002-7870-2219}\refaddr{label1,label2}, Yu. Kalyuzhnyi\orcid{0000-0002-0631-9982}\refaddr{label1,label3},
        O. Patsahan\orcid{0000-0002-5839-3893}\refaddr{label1},
        M.~Holovko\orcid{0000-0001-8114-5356}\refaddr{label1}
    }
\addresses{
\addr{label1} Institute for Condensed Matter Physics of the National
Academy of Sciences of Ukraine, 1 Svientsitskii Str., 79011 Lviv,
Ukraine
\addr{label2} Institute of Applied Mathematics and Fundamental Sciences, Lviv Polytechnic National University, 12 S.~Bandera Str., 79013 Lviv, Ukraine
\addr{label3} University of Ljubljana, Faculty of Chemistry and Chemical Technology,  Ve\v{c}na pot 113, SI-1000
Ljubljana, Slovenia
}
\date{Received April 17, 2024, in final form June 12, 2024}

\begin{document}

\maketitle

\begin{abstract}
We  develop a theory for the description 
of  ionic liquids (ILs) confined in a porous medium formed by a matrix of immobile randomly placed uncharged  particles. The IL is modelled as an electroneutral  
mixture of hard-sphere anions and flexible linear chain cations, represented by tangentially bonded hard spheres with the charge located on one of the terminal beads. The theory combines  a generalization of the scaled particle theory, Wertheim's thermodynamic perturbation theory, and the associative mean-spherical approximation  and allows one to obtain analytical expressions for the
pressure and chemical potentials of the  matrix--IL system.
Using the theory, we calculate the vapour–liquid phase diagrams for two versions of the IL model, i.e., when
the cation is modelled as a dimer and as a chain, in a complete association limit.
The effects of the matrix confinement  and of the  non-spherical shape of the cations on the vapour-liquid phase diagrams are studied.
%
%
\printkeywords

\end{abstract}

\section{Introduction}

We dedicate the article to the memory of Ihor Yukhnovsky, who recently passed away, and who, starting from the 1950s, made a significant contribution to the development of the theory of electrolyte solutions. He founded a scientific school on the theory of ionic systems in Lviv, of which the authors of this article have the honor to be a part.

Ionic liquids (ILs) are a subclass of molten salts with a melting temperature  lower than 373~K. 
In general, ILs are composed
of organic cations with either inorganic or organic anions. There are many
possibilities of making ILs due to the variety of ions
and the variations in side chains of the ions.
ILs and ionic solutions of complex molecular ions, due to their unique
physical and chemical properties, play an important role in fundamental research and have an important applied value in modern technological processes  \cite{Buzzeo2004, Silvester2006,Suo2015,Correia2020,Egorova2017}. 
A significant part of technological applications is based on the properties of ionic liquids in porous materials. In particular, ILs in porous electrodes are potentially important in innovative electrochemistry \cite{singh2014ionic,Kondrat2023}. The presence of a porous medium can substantially affect the  properties of ILs including their  phase behavior. 
The development of a theory for the study and prediction of the phase behavior of ILs with a complex structure of ions under disordered porous confinement still remains a relevant problem.

Phase behavior of the bulk ionic solutions with  Coulomb-dominated interactions has been a subject of active research  for the last few decades. 
The simplest and
most frequently used model for such systems is a restricted primitive model (RPM) which consists of equal numbers of equisized positively and negatively charged
hard spheres (HSs) immersed in a structureless dielectric continuum. The RPM undergoes a vapour-liquid-like phase transition at low
temperature and at low density (see review~\cite{patsahan2012order} and the literature cited therein).
Further studies  focused on the modified versions of the RPM that takes into account charge and/or size asymmetry of ions. 

{ Significant progress in a theoretical description of the vapour-liquid phase behavior in the above-mentioned models was made within the approaches that take into account ion association. Among them there is a generalized Debye-H\"uckel theory proposed by Fisher and Levin \cite{Fisher_Levin_1993,levin1996criticality}, the associative mean spherical approximation  (AMSA) \cite{holovko1991effects,KALYUZHNYI1998,Kalyuzhnyi2000}, and the binding mean spherical approximation (BIMSA)~\cite{Blum95,Bernard96,jiang2001charged,Jiang02,Qin2004}. It is worth noting that both the AMSA and the BIMSA are, in fact, identical.  In addition, the method of collective variables (CVs) appeared to be successful for the description of these models~\cite{Caillol2005,Patsahan2006,Patsahan2010}.  The advantage of this approach is that one can derive an analytical expression for the relevant chemical potential which includes the effects of higher-order correlations between the ions. The historical discussion of the vapour-liquid phase behaviour and criticality in the RPM is available in \cite{patsahan2012order,holovko2017effects}.}

Despite a significant progress in understanding  the phase behavior of relatively simple models of electrolyte solutions, today's challenges require consideration of more complex ionic systems. 
Real ILs and ionic solutions are characterized by a specific intermolecular interaction that is important at both short and long distances. In particular, all these objects, in addition to the presence of charges, have  a complex molecular structure and  geometry of ions which are far from spherical shape.  The location of charges  is also important, i.e., the charge is frequently not located at the center of mass of molecular ions.
These features can have a significant impact on the properties of ILs. 
This also  complicates the development of theoretical approaches capable of properly predicting the properties of these systems. 

Several simple IL models have been proposed  over the past fifteen years
\cite{Malvaldi2007,Spohr2009,MartnBetancourt2009,Fedorov2010,Wu2011,
	Ganzenmller2011,Lindenberg2014,Lindenberg2015,Guzmn2015,SilvestreAlcantara2016,Lu2016,Kalyuzhnyi2018}. Molecular ions  within the framework of these models are represented either as  hard spheres with off-center point charges~\cite{Spohr2009,Ganzenmller2011,Lindenberg2014,SilvestreAlcantara2016}, as dimers with point charges located on one or both ends~\cite{Malvaldi2007,Fedorov2010,Wu2011, Ganzenmller2011,Kalyuzhnyi2018}, or as  hard spherocylinders with a point charge located on one of their ends \cite{MartnBetancourt2009}.
These models  are mostly studied using computer simulation methods.	
The vapour-liquid phase behavior of  model ILs with chain-like molecular ions was studied theoretically in \cite{Guzmn2015,Kalyuzhnyi2018}. However, these studies were limited to the bulk case.

The  most frequently used model  for the study of fluids in disordered porous media is the model proposed by Madden and Glandt~\cite{madden1988distribution}. 
In this model, the porous medium is  represented 
as a quenched disordered
matrix of HSs while the fluid is distributed inside
the matrix. In this case, statistical-mechanical averages used for calculations of thermodynamic properties become double ensemble averages: the first average is taken over all degrees of freedom of fluid particles  keeping the quenched  particles fixed, and the other average is performed  over all realizations of a matrix. 
Using this approach, the scale particle theory (SPT) was developed and analytical expressions for  the thermodynamic functions of the HS fluid in a disordered HS matrix were obtained
\cite{holovko2009highly,patsahan2011fluids,holovko2012fluids,holovko2012one,holovko2017improvement}.
This allowed one to use the model of a HS fluid in a disordered HS matrix as a reference system (RS). Therefore, to study the effect of the presence of a porous medium  on the phase behavior in the RPM fluid, both the CV method~\cite{holovko2016vapour}  and the AMSA approach~\cite{ holovko2017effects}  were used.
It was shown that the presence of a porous medium results in a shift of the vapour-liquid phase diagram of the RPM towards lower densities and temperatures.
Later on SPT was generalized and applied to the study of the properties of the multicomponent HS fluid confined in the
HS disordered matrix~\cite{chen2016scaled}. This version of the SPT was used to extend the theory proposed
in \cite{holovko2016vapour,holovko2017effects} for the case  of primitive models with ions of a spherical shape and 
different sizes ~\cite{holovko2017application,patsahan2018vapor}, as well as  the models involving oppositely charged spherical ions differing in both size and valence~\cite{HolPatPat18charge}. 

In this work, we study the vapour-liquid phase diagrams of ILs in a disordered porous medium. 
Following \cite{Kalyuzhnyi2018}, we  model an IL as a two-component mixture of
HS anions and flexible linear chain cations, represented by
the tangentially bonded HSs with the charge located on one
of the terminal beads.
The existing theoretical approaches used for describing the chain ions provide a fairly good description of the thermodynamics of these systems in the bulk case. However, in the presence of a disordered matrix, this problem is much more difficult and has remained unsolved until recently.
We propose a theoretical approach that  allows one 
to obtain the thermodynamic functions of the system of chain ions
in a disordered porous matrix. The approach combines the extension of the AMSA theory for chain-forming fluids \cite{Wertheim1984,Wertheim1984-2,Wertheim1986,Wertheim1986-2,holovko1991effects}
and the extension of the SPT theory for HS fluids in a disordered HS matrix~\cite{reiss1959statistical,reiss1960aspects,lebowitz1965scaled, boublik1974statistical, patsahan2011fluids,holovko2017improvement}.
Using this approach we  calculate the vapour-liquid phase diagrams  of the   IL models with molecular ions, represented by homo- and hetero-nuclear charged chains confined in a disordered HS matrix of different porosities.

The remainder of the paper is organised as follows.	 In section~\ref{sec2}, we present the models and develop the theoretical formalism. The results
are presented and discussed in section~\ref{sec3}. We conclude in section~\ref{sec4}.

\section{Models and theory }	\label{sec2}

\subsection{Models}

We present an IL as an electroneutral mixture of cations and anions with the number densities  $\rho_{c}$ and $\rho_{a}$ ($\rho_i=N_i/V$), respectively, immersed in a structureless dielectric continuum. The anions are modelled as charged HSs with diameter $\sigma_1=\sigma_-$ and charge $ez_{1}=ez_{-}$.
The cations are represented  by $m-1$ tangentially
bonded HS monomers with the charge located on one of the terminal beads.
The charged monomer of the cation has a diameter  $\sigma_2=\sigma_+$ and a charge  $ez_2=ez_{+}$, while all uncharged monomers have the same diameter  $\sigma_n$. It is assumed that the charges are in the center of the spheres.

\begin{figure}[h]
	\begin{center}
		\includegraphics[width=0.6\textwidth]{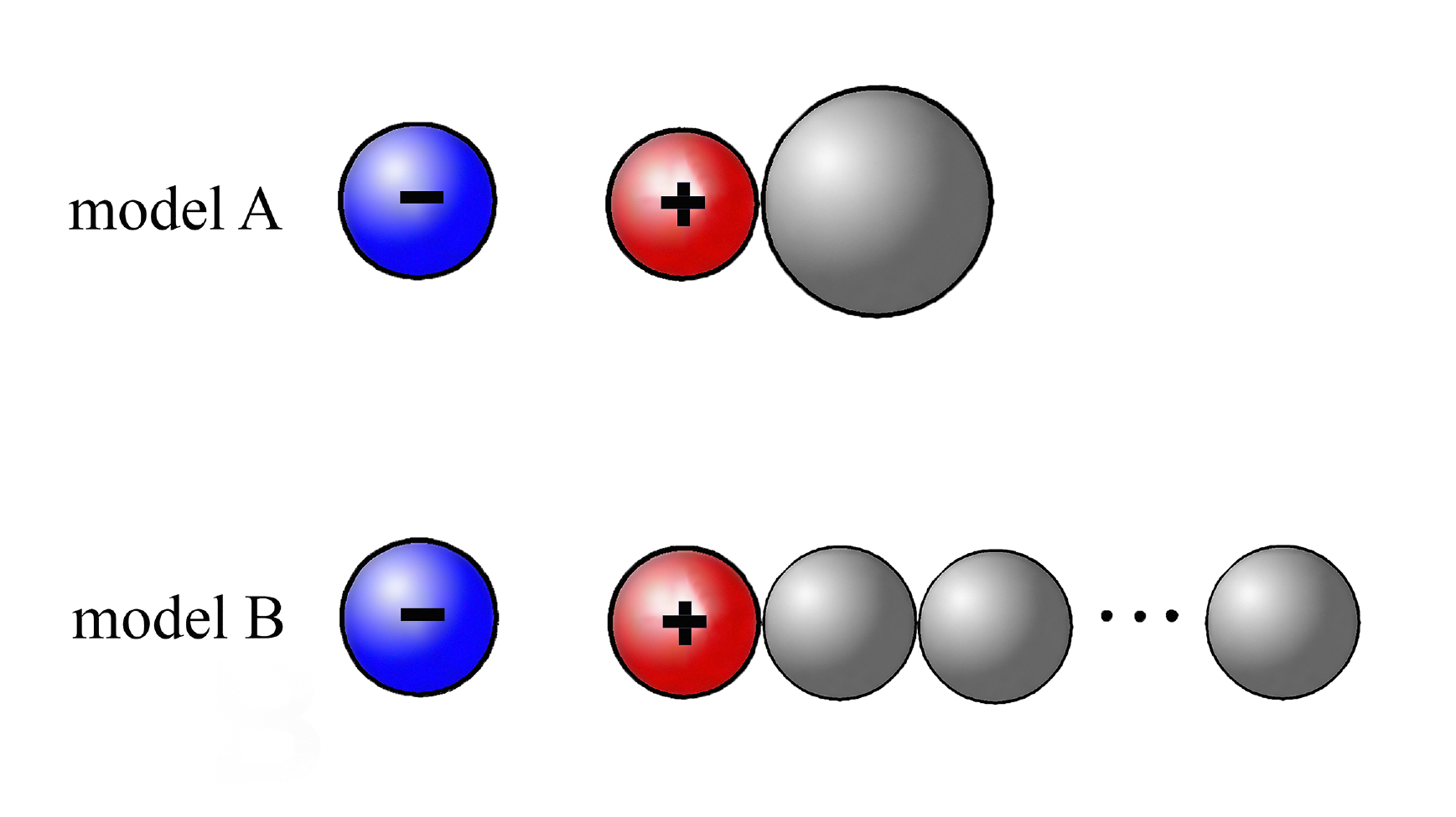}
		\caption{(Colour online) Two models of  ILs distinguished by the model for the cation: hetero-nuclear dimer (model~A) and mono-nuclear chain (model~B). In both cases, the anion is modelled as a single charged hard sphere.}
		\label{models}
	\end{center}
\end{figure}

The pair potential acting between the particles is
represented by the sum of site-site HS potentials $U_{ij}^{\text{hs}}(r)$:
\begin{eqnarray}
	U_{ij}^{\text{hs}}(r) = \left\{
	\begin{array}{ll}
		\infty, & r<\sigma_{ij}=(\sigma_{i}+\sigma_{j})/2,\\
		0,& r\geqslant \sigma_{ij}
	\end{array}
	\right. 
	\label{u_hs}
\end{eqnarray}
and Coulomb potential $U_{ij}^{\text{C}}(r)$:
\begin{equation}
	U_{ij}^{\text{C}}(r)=\frac{e^2 z_{i}z_{j}}{\varepsilon r},
	\label{u_C}
\end{equation}
between charged HSs. Here, $\varepsilon$ is the dielectric constant of the continuum. In general, we assume that
\[
\sigma_{+}=\sigma_{-}=\sigma\neq\sigma_{n}
\]
and  $|z_1|=|z_2|=z$. The total number density of the system is  $\rho_{t}=2\rho$, where $\rho=\rho_{-}=\rho_{+}$. For   $\sigma_{+}=\sigma_{-}=\sigma=\sigma_{n}$, the cation model reduces to a homo-nuclear chain.  
Two versions of the above described model of ILs are presented in figure~\ref{models}.

The IL model is confined in a disordered porous matrix formed
by hard spheres of diameter $\sigma_0$. The interaction potentials between an ion
and a matrix particle $U_{0i}^{\text{hs}}(r)$ ($i=+,-,n$) and between two matrix particles $U_{00}^{\text{hs}}(r)$ are described by the HS potentials:
\begin{eqnarray}
	U_{0i}^{\text{hs}}(r) = \left\{
	\begin{array}{ll}
		\infty, & r<\sigma_{0i}\\
		0,& r\geqslant \sigma_{0i}
	\end{array}
	\right. \,,           \qquad
	U_{00}^{\text{hs}}(r) = \left\{
	\begin{array}{ll}
		\infty, & r<\sigma_{0}\\
		0,& r\geqslant \sigma_{0}
	\end{array}
	\right. .
	\label{ion_matrix}
\end{eqnarray}
In (\ref{ion_matrix}), $\sigma_{0i}=\frac{1}{2}\left(\sigma_{0}+\sigma_{i}\right)$
and, in general,  $\sigma\neq\sigma_{n}\neq\sigma_{0}$. 

The matrix is characterised by different types of porosity, namely, geometrical porosity  $\phi _{0}$ and  probe-particle porosity $\phi_i$  for the $i$th  species of the HS mixture. The probe-particle porosity $\phi_i$  is defined by the excess value of the chemical potential of a fluid particle with diameter  $\sigma_i$  in
the limit of infinite dilution and, hence, takes into account the size
of adsorbate species \cite{chen2016scaled}. Explicit expressions for $\phi _{0}$  and $\phi_i$  will be given below. In order to
distinguish a pure geometrical effect of confinement, we restrict the
model only to a hard-core interaction between the fluid and matrix
particles  with the latter being uncharged.

We present  Helmholtz free energy $f=F/V$ of the matrix-IL
system [equations~(\ref{u_hs})--(\ref{ion_matrix})] in the form:
\[
\beta f=\frac{\beta F}{V}=\beta f^{(\text{id})}+\beta f^{(\text{ref})}+\beta\Delta f,
\]
where $\beta f^{\text{id}}$ is an ideal-gas contribution, $f^{\text{ref}}$ is the free energy of a reference system (RS), 
which is represented by a multi-component HS fluid confined in the HS matrix.  $\Delta f$ is the part 
connected with the ionic subsystem, and $\beta =1/k_{\text{B}}T$.  Below we {present} expressions for
these contributions to the free 
energy  and the corresponding contributions to the pressure and chemical potentials. {Here, we provide a brief description of each theoretical approach upon which our theory is based. A detailed presentation of the generalized SPT theory and the extension of the AMSA theory for chain-forming fluids can be found elsewhere \cite{holovko2015physics,holovko2017improvement,chen2016scaled, holovko1991effects, Protsykevytch1997,Bernard2000,Kalyuzhnyi2001,holovko2005concept}.}

\subsection{Reference system}

To calculate thermodynamic properties of the RS we use a generalized version of
the SPT theory, in particular, the  approach, proposed recently for a two-component mixture of hard
convex bodies confined in a disordered HS matrix~\cite{Hvozd2022}. In order to
characterize each particle of the system, we use three geometrical
parameters:  the volume $V$ of a particle, its surface area $S$  and the
mean curvature $r$ taken with a factor ${1}/{4\piup}$~\cite{holovko2015physics}.
For a multicomponent HS mixture  and for HS matrix particles we have:
\begin{equation}
	\label{funct1}
	V_{i}=\frac{1}{6}\piup \sigma_{i}^3\;,\;\;\;   S_{i}=\piup \sigma_{i}^2\;,\;\;\;   r_{i}=\sigma_{i}/2 \;,\; \; i=0,\ldots, m.
\end{equation}


The geometrical
porosity depends on a matrix structure and is related to
the volume of a void existing between the matrix particles, i.e.,
\begin{equation}
	\label{phi0}
	\phi_0=1-\eta_0,
\end{equation}
where  $\eta_{0}=\piup\rho_{0}\sigma_{0}^{3}/6$ is the packing fraction of the matrix particles. 
 {The second type of porosity is defined by the chemical potential of a fluid in the limit of infinite dilution and it is called a
 probe-particle porosity~$\phi$. This porosity characterises the adsorption of a fluid particle in an empty matrix. For a multicomponent
HS fluid confined in a hard-sphere matrix,
 we have the probe-particle porosity $\phi_i$ for each species $i$.} The probe-particle porosity $\phi$ depends only on the nature of the $i$th species and has the form~\cite{chen2016scaled}:
\begin{eqnarray*}
	\phi_{i}&=&(1-\eta_0)\exp \bigg[-3k_{i0}\left(1+k_{i 0}\right)\frac{\eta_0}{1-\eta_0}-\frac{9}{2} k_{i0}^2 \frac{\eta_0^2}{(1-\eta_0)^2} \nonumber \\
	&-&k_{i 0}^3 \frac{\eta_0}{(1-\eta_0)^3}\left(1+\eta_0+\eta_0^2\right)\bigg],
\end{eqnarray*} 
where $k_{i0}=\sigma_i/\sigma_0$.

We present the pressure of the HS fluid mixture confined in a disordered HS matrix in the form \cite{Hvozd2022}:
\begin{eqnarray}
	\label{pressureCS1}
	\beta P^{\rm{ref}}=\beta P^{\rm{SPT2b3}^\star}+\beta \Delta P^{\rm{CS}},
\end{eqnarray}
where  $\beta P^{\rm{SPT2b3}^\star}$ is the contribution  due to the  SPT2b3* approximation \cite{holovko2017improvement}.
This approximation provides  a correct description of the thermodynamic properties of the RS in a wide
range of a fluid packing fraction $\eta=\piup\sum_{i=1}^m\rho_{i}\sigma_{i}^3/6$,  starting from the smallest ones and ending with the packing fractions close to the maximum values possible in a given matrix. $ \beta\Delta P^{\rm{CS}}$  is the Carnahan-Starling correction for the case of a mixture in a porous
medium \cite{boublik1974statistical}.
Following  \cite{Hvozd2022}, we obtain:
\begin{eqnarray}
	\label{pressureSPT2b3star}
	\frac{\beta P^{\rm{SPT2b3}^\star}}{\rho_m}&=&
	\frac{1}{1-\eta/\phi_0}
	+\frac{A}{2}\frac{\eta/\phi_0}{\left(1-\eta/\phi_0\right)^2}
	+\frac{2B}{3}\frac{\left(\eta/\phi_0\right)^2}{\left(1-\eta/\phi_0\right)^3}
	\nonumber \\
	&
	+&\frac{\phi_0-\phi^\star}{\phi^\star}\frac{\phi_0}{\eta}\left[\ln(1-\eta/\phi_0)+\frac{\eta/\phi_0}{1-\eta/\phi_0}\right] \nonumber \\
	&+&\frac{\phi^\star-\phi}{\eta}\left[\ln(1-\eta/\phi^\star)+\frac{\eta/\phi^\star}{1-\eta/\phi^\star}\right],
\end{eqnarray}
\begin{eqnarray}
	\label{pressCS1}
	\frac{\beta \Delta P^{\rm{CS}}}{\rho_m}=-\frac{\left(\eta/\phi_0\right)^3}{\left(1-\eta/\phi_0\right)^3}\frac{q_m s_m^2}{9 v_m^2},
\end{eqnarray}
where $\phi^\star=\frac{\phi_0\phi}{\phi_0-\phi}\ln\frac{\phi_0}{\phi}$ 
is  the maximum value of the packing fraction of the fluid in the disordered matrix~\cite{holovko2012one}, $\phi$ is the total thermodynamic porosity of the matrix for a given fluid mixture:
\begin{equation}
	\label{smallphip}
	\frac{1}{\phi}=\frac{1}{\eta} \sum_{i=1}^{m} \frac{\rho_i V_i}{\phi_i}. 
\end{equation}
$A$ and $B$ are given by
\begin{equation}
	\label{Ap}
	A=\sum_{i=1}^{m}x_i a_i, \qquad
	B=\sum_{i=1}^{m}x_i b_i,
\end{equation}
where the coefficients $a_i$ and $b_i$  define the porous medium structure \cite{chen2016scaled,Hvozd2022}, the expressions for them are too cumbersome to be presented here  (they can be found in \cite{Hvozd2022}),
$x_i=\rho_i/\rho_m$ denotes the fluid composition, and $v_m$, $s_m$, $q_m$, and $\rho_m$ have  the form \cite{boublik1974statistical}:
\begin{eqnarray}
	v_m=\sum_{i=1}^m x_i V_i\,,\,\,\,\,s_m=\sum_{i=1}^m x_i S_i\,,\,\,\,\, 
	q_m=\sum_{i=1}^m x_i r_i^2\,, \,\,\, \rho_m=\sum_{i=1}^m \rho_i, 
	\label{v-s-q}	
\end{eqnarray} 
where $V_i$, $S_i$, and $r_i$ are given in (\ref{funct1}), $\rho_i$ is the number density of the $i$th species of the HS mixture.

Similarly, we present 
the partial chemical potential $\beta \mu_{i}$  as follows: 
\begin{eqnarray}
	\label{mu_ref}
	\beta\mu_i^{(\text{ref})}=\beta\mu_i^{\rm{SPT2b3^*}}+\beta\Delta\mu_i^{\rm{CS}},
\end{eqnarray}
\begin{eqnarray}
	\label{chemSPT2b3star}
	\beta \mu_{i}^{\rm{SPT2b3}^\star}&=&\beta \mu_{i}^{\rm{SPT2a}}
	+\frac{\eta(\phi_0-\phi^\star)}{\phi_0\phi^\star\left(1-\eta/\phi_0\right)}
	+\frac{\eta\left(\phi^\star-\phi\right)}{\phi^\star\phi^\star\left(1-\eta/\phi^\star\right)} \nonumber\\
	&+&\left(\frac{\rho V_i}{\eta}-1\right)\bigg[\frac{\phi_0-\phi}{\eta}\ln(1-\eta/\phi_0)+\frac{\phi(\phi_0-\phi^\star)}{\phi_0\phi^\star\left(1-\eta/\phi_0\right)} \nonumber\\
	&+&\frac{\phi\left(\phi^\star-\phi\right)}{\phi^\star\phi^\star\left(1-\eta/\phi^\star\right)}\bigg]-\frac{\rho V_i}{\eta}\left(\frac{\phi}{\phi_i}-1\right)\bigg[\frac{\phi}{\eta}\ln\left(1-\eta/\phi_0\right) \nonumber\\
	&-&\frac{\phi(\phi_0-\phi^\star)}{\phi_0\phi^\star\left(1-\eta/\phi_0\right)}
	-\frac{\phi\left(\phi^\star-\phi\right)}{\phi^\star\phi^\star\left(1-\eta/\phi^\star\right)}+1\bigg],
\end{eqnarray}
\begin{eqnarray}
	\label{muCS}
	\beta\Delta\mu_i^{\rm{CS}}&=&-\frac{V_i}{v_m}\frac{\left(\eta/\phi_0\right)^3}{\left(1-\eta/\phi_0\right)^3}\frac{q_m s_m^2}{9 v_m^2}
	+\frac{s_m}{9v_m^3} \big[\left(q_i s_m +2S_i q_m \right)v_m -2V_i q_m s_m\big] \nonumber \\ 
	&\times&\bigg[\ln(1-\eta/\phi_0)+\frac{\eta/\phi_0}{1-\eta/\phi_0}-\frac{1}{2}\frac{\left(\eta/\phi_0\right)^2}{\left(1-\eta/\phi_0\right)^2}\bigg],
\end{eqnarray}
where  $\beta \mu_{i}^{\rm{SPT2a}}$ is given in \cite{Hvozd2018}. The other notations are the same as in (\ref{pressureSPT2b3star})--(\ref{pressCS1}).

\subsection{Ionic subsystem}
We consider an ionic subsystem consisting of an electroneutral mixture of chain cations and monomeric anions in a structureless dielectric medium. The model can be considered as a partial case of the $m$-component HS mixture with two ``sticky'' spots (patches) of the type $A$ and $B$, randomly placed
on the surface of each particle forming the cation chain, simultaneously assuming an infinitely strong attraction between
the patches of the type $B$ and $A$ located on the surface of the particles
of the type $i$ and $i + 1$, respectively. Importantly, the size of each patch is small enough to ensure that only one $A$-$B$ bond  can be created.
Thermodynamic properties for such a generalized model were obtained using a modified Wertheim's multidensity  Ornstein-Zernike (OZ)  equation supplemented by the AMSA  theory~\cite{holovko1991effects} formulated for the case of chain-forming fluids  \cite{Protsykevytch1997,Kalyuzhnyi2001}. Here, we present only the final expressions for the IL model under consideration.  

The solution of the AMSA theory for our IL model can be reduced to the
solution of a nonlinear algebraic equation for Blum's screening parameter $\Gamma$ \cite{Protsykevytch1997,Bernard2000,Kalyuzhnyi2001,Kalyuzhnyi2018}:
\begin{equation}
	\label{Gamma-2}
	\Gamma^2=\frac{\piup\beta e^2}{\varepsilon}\rho\sum_{i=1}^{m}\mathbf{X}_{i}\mathbf{\alpha}\mathbf{X}_{i}^T,
\end{equation}
where $\mathbf{X}_{i}=(X_{i}^0,X_{i}^A,X_{i}^B)$ is a row matrix and $\mathbf{\alpha}$ is a symmetrical $3\times 3$ matrix
\begin{eqnarray*}
	\mathbf{\alpha}=
\begin{pmatrix}
		1 & 1 & 1 \\
		1 & 0 & 1 \\
		1 & 1 & 0
	\end{pmatrix}.
\end{eqnarray*}
A general form of the matrix elements ${X}_{i}^{\alpha}$ with $i=1,2,\ldots,m$ and $\alpha=0,A,B$ was obtained in  \cite{Kalyuzhnyi2018}.

As will be shown below, the elements of the matrix $\mathbf{X}_{i}$ contain the quantity $t$ 
\begin{equation}
	\label{t}
	t=2\piup\sigma_{12}^2x^2K_{\text{as}}^{(0)}\exp\left(G_{00}(\sigma_{12}^+)-\beta U^{(C)}(\sigma_{12}^+)\right)g_{12}^{00}(\sigma_{12}^+)|_{z_{i}=0},
\end{equation}
where  $g_{12}^{00}(\sigma_{12}^{+})|_{z_{i}=0}$ is the contact value of the radial distribution function  at zero charges on the 
anion and cation bead, $
G_{00}(\sigma_{12}^+)=g_{12}^{00}(\sigma_{12}^+)-g_{12}^{00}(\sigma_{12}^+)|_{z_{i}=0}$,
$K_{\text{as}}^{(0)}$ is the association constant, $x$ is the fraction of free anions
(or cations)  determined from the solution of the mass action 
law type of equation
\begin{equation}
	\label{eq-x}
	2\rho t\sigma_{12}+x-1=0.
\end{equation}
In order to obtain (\ref{t}), the exponential approximation \cite{holovko2005concept} is used.

We consider two versions (model~A and model~B) of the IL model  which differ by the cation shape  (see figure~\ref{models}). In model~A, the cations are modelled as a hetero-nuclear dimers ($\sigma_-=\sigma_+=\sigma\neq\sigma_n$), where indices  $-,+,n$ indicate anion, cation, and neutral monomer, respectively) and in model~B the cations are modelled as homo-nuclear chains  ($\sigma_1=\sigma_2=\ldots=\sigma_m=\sigma$). 
In both models, the anions are modelled as single charged HSs. The expressions for ${X}_{i}^{\alpha}$ for each of the models are presented below.

\paragraph{{Model~A: the cation modelled as hetero-nuclear dimer.}}
In this case, the matrix elements $\mathbf{X}_{i}$ have the form:\\
for $\mathbf{X}_{1}=\mathbf{X}_{-}$,
\begin{eqnarray}
	X_{-}^0&=&-\Gamma_{\sigma_{-}}(z+\eta^B\sigma_{-}^2), \qquad X_{-}^A=0,\label{Xa0}   \\
	X_{-}^B&=&\Gamma_{\sigma_{-}}\Gamma_{\sigma_{+}}\sigma_{-}\rho t\bigg(z-\eta^B\sigma_{+}^2-\eta^B\Gamma_{\sigma_{n}}\frac{\sigma_{+}\sigma_{n}^2}{2\sigma_{+n}}   \bigg), \nonumber
\end{eqnarray}
for $\mathbf{X}_{2}=\mathbf{X}_{+}$,
\begin{eqnarray}
	X_{+}^0&=&\Gamma_{\sigma_{+}}(z-\eta^B\sigma_{+}^2),\label{Xc0} \\ X_{+}^A&=&-\Gamma_{\sigma_{-}}\Gamma_{\sigma_{+}}\sigma_{+}\rho t(z+\eta^B\sigma_{-}^2), \nonumber\\
	X_{+}^B&=&-\Gamma_{\sigma_{+}}\Gamma_{\sigma_{n}}\eta^B\frac{\sigma_{+}\sigma_{n}^2}{2\sigma_{+n}}, \nonumber
\end{eqnarray}
for  $\mathbf{X}_{3}=\mathbf{X}_{n}$,
\begin{eqnarray*}
	X_{n}^0&=&-\Gamma_{\sigma_{n}}\eta^B\sigma_{n}^2, 	
	\\ X_{n}^A&=&\frac{\Gamma_{\sigma_{+}}\Gamma_{\sigma_{n}}\sigma_{n}}{2\sigma_{+n}}\left[z-\eta^B\sigma_{+}^2-\rho t\Gamma_{\sigma_{-}}\sigma_{+}(z+\eta^B\sigma_{-}^2)\right],
	\\
	X_{n}^B&=&0.
\end{eqnarray*}
The expression for $\eta^B$ is as follows:
\begin{equation}
	\eta^B={{\piup\rho\over 2\Delta} \left\{\sigma_+\Gamma_{\sigma_+}-\sigma_{-}\Gamma_{\sigma_{-}}+
		\left[
		{\sigma_n^2\Gamma_{\sigma_n}\over 2\sigma_{n+}}+
		\rho\Gamma_{\sigma_{-}}t\left(\sigma_{-}^2-\sigma_+^2-{\sigma_+\sigma_n^2\Gamma_{\sigma_n}\over 2\sigma_{n+}}\right)\right]\Gamma_{\sigma_+}\right\}\over 
		1+{\piup\rho\over 2\Delta}\left\{\sum_{i=-}^n\sigma_i^3\Gamma_{\sigma_i}
		+\sigma_+^2\left[{\sigma_n^2\Gamma_{\sigma_n}\over  2\sigma_{n+}}+
		\sigma_{-}^2\rho\Gamma_{\sigma_{-}}t
		\left(2+{\sigma_n^2\Gamma_{\sigma_n}\over \sigma_{+n}\sigma_+}\right)
		\right]\Gamma_{\sigma_+}\right\}},
	\label{eta-B}
\end{equation}
where
$\sigma_{+n}=(\sigma_++\sigma_n)/2$,  $\Delta$ and $\Gamma_{\sigma_i}$ have the form:
\begin{equation}
	\Delta=1-\piup\rho\sum_{i=1,\ldots,m}\sigma_{i}^3/6,
	\label{Delta}
\end{equation}
\begin{equation}
	\Gamma_{\sigma_i}=\left(1+\sigma_i\Gamma\right)^{-1}.
	\label{Gamma-sigmai}	
\end{equation}

\paragraph{{Model~B:  a chain cation with monomers of the same size.}}
In the case where $m$ monomers forming a chain cation are HSs of the same diameter,
the elements ${X}_{i}^{\alpha}$ of the matrix $\mathbf{X}_{i}$ can be presented in the following compact form \cite{Kalyuzhnyi2018}:\\
\begin{eqnarray}
	X_i^0&=&\left[z_i-\eta^B\sigma_i^2\right]\Gamma_{\sigma_i}, 
	\label{X0} \\
	X_i^\alpha&=&\sigma_i\left[\tau_i^\alpha(z)-\eta^B\tau_i^\alpha(\sigma^2)\right],
	\;\;\;\;\;\;\alpha\neq 0\;\;(\alpha=A,B),
	\nonumber
	\\
	\eta^B&=&{{\piup\over 2\Delta}\rho\sum_{i=1}^m\sigma_i\left\{z_i\Gamma_{\sigma_i}+\sigma_i\left[
		\tau_i^A(z)+\tau_i^B(z)\right]\right\}\over
		1+{\piup\over 2\Delta}\rho\sum_{i=1}^m\sigma_i^2\left[\sigma_i\Gamma_{\sigma_i}
		+\tau^A_i(\sigma^2)+\tau_i^B(\sigma^2)\right]},
	\nonumber
\end{eqnarray}
where the following notations are introduced:
\[
	\tau^A_1(y)=0, \quad \tau_2^A(y)=\rho\Gamma_{\sigma_{-}}\Gamma_{\sigma_+}y_{-}t,
\]
	\[
	\tau_3^A(y)={\Gamma_{\sigma_n}\Gamma_{\sigma_+}\over 2\sigma_{n+}}\left(
	y_{+}+\rho\sigma_+y_{-}\Gamma_{\sigma_{-}}t\right),
\]
\[
	\tau_i^A(y)={\Gamma_{\sigma_n}^2\over 2\sigma_{n}}\left[\left(
	y_{+}+\rho\sigma_+y_{-}\Gamma_{\sigma_{-}}t\right){\sigma_n\Gamma_{\sigma_+}\over 2\sigma_{n+}}
	\left({\Gamma_{\sigma_n}\over 2}\right)^{i-4} 
	+y_n\sum_{l=4}^i
	\left({\Gamma_{\sigma_n}\over 2}\right)^{i-l}\right], 
	4\leqslant i\leqslant m,
\]
\[	
	\tau_1^B(y)=\rho\Gamma_{\sigma_{-}}\Gamma_{\sigma_+}t\left[y_{+}+y_n{\sigma_+\over\sigma_{+n}}
	\sum_{l=3}^m\left({\Gamma_{\sigma_n}\over 2}\right)^{l-2}\right],
	\]
	\[
	\tau_2^B(y)={\Gamma_{\sigma_+}\Gamma_{\sigma_n}\over 2\sigma_{+n}}y_n
	\sum_{l=3}^m\left({\Gamma_{\sigma_n}\over 2}\right)^{l-3},
	\]
\[
	\tau_i^B(y)={\Gamma_{\sigma_n}^2\over 2\sigma_n}y_n
	\sum_{l=i+1}^m\left({\Gamma_{\sigma_n}\over 2}\right)^{l-i-1},
	\;\;\;\;\;3\leqslant i<m, \quad \tau_m^B(y)=0,
\]
and $y$  is taking on the values either $z$ or $\sigma^2$.   $\Gamma_{\sigma_i}$ and $\Delta$ entering  the above equations are presented in  (\ref{Delta}) and (\ref{Gamma-sigmai}), respectively. 

\subsection{Structural properties}

Equations (\ref{t})--(\ref{eq-x}) include contact values of the radial distribution function $g_{+-}^{00}(\sigma_{+-})$ between oppositely charged ions of the matrix--IL model, which includes charges and the model with $z_{i}=0$, respectively \cite{Protsykevytch1997,Kalyuzhnyi2001,Kalyuzhnyi2018}
\begin{equation}
	\label{contact_full}
	g_{+-}^{00}(\sigma_{+-})=g_{+-}^{00}(\sigma_{+-})|_{z_{i}=0}-\frac{\beta e^2}{\varepsilon\sigma_{ij}}X_{+}^{0}X_{-}^0,
\end{equation}
where the expressions for $X_{+}^0$ and $X_{-}^0$ are the same for both models (see (\ref{Xa0}), (\ref{Xc0}), and (\ref{X0})). 

The contact value of the radial distribution function of the model at zero charge within the framework of the AMSA theory coincides with the contact value of the corresponding radial distribution function of the $m$-component HS mixture. For our model, this will be the contact value of the radial distribution function of a two-component HS mixture  in the presence of immobile matrix particles. It can be obtained by generalizing the result of \cite{Kalyuzhnyi:2014,holovko2020korvatska} for the case of a two-component HS fluid in a disordered porous medium. According to the SPT, the contact value of the radial distribution function between a scaled particle of infinitesimally small size $\sigma_{s}$ and the $i$th species particle of HS mixture confined in a porous matrix can be presented as follows \cite{holovko2009highly,chen2016scaled}:
\begin{displaymath}
	g_{i s}^{\text{hs}}(\sigma_{i s})=\frac{1}{p_{0}^{i}(\sigma_{s})-\sum_{b}\eta_{b}(1+\sigma_{s}/\sigma_{i})^{3}},
\end{displaymath}
where $\sigma_{\alpha s}=(\sigma_{\alpha}+\sigma_{s})/2$, $p_{0}^{\alpha}(\sigma_{s})$ is the probability of finding a cavity created by the scaled particle in the matrix in the absence of fluid particles \cite{holovko2012fluids}. For a point scaled particle, $p_{0}^{\alpha}(\sigma_{s}=0)$ is  the geometric porosity of the matrix $\phi_{0}$ \cite{chen2016scaled}. To obtain the contact value of the radial distribution function  between two fluid particles of type $i$ and $j$, we follow  \cite {Kalyuzhnyi:2014,Hvozd2022} and 
express $g_{i s}^{\text{hs}}(\sigma_{i s})$ as
\begin{displaymath}
	g_{i s}^{\text{hs}}(\sigma_{i s})=G_{i s}^{(0)}+G_{i s}^{(1)}\frac{\sigma_{s}}{\sigma_{s}+\sigma_{i}}+\frac{1}{2}G_{i s}^{(2)}\left(\frac{\sigma_{s}}{\sigma_{s}+\sigma_{i}}\right)^2,
\end{displaymath}
where  $G_{i s}^{(0)}$,  $G_{i s}^{(1)}$ and $G_{i s}^{(2)}$ are found  from the continuity of $g_{i s}^{\text{hs}}(\sigma_{i s}^{+})$ and the first and second derivatives with respect to $\sigma_{s}$ at $\sigma_{s}=0$. 

As a result,  we get for $g_{+-}^{(\text{hs})}(\sigma_{+-})$
\begin{eqnarray}
	\label{g12}
	g_{+-}^{(\text{hs})}(\sigma_{+-})&=&\frac{1}{\phi_0-\eta}+\frac{3}{2}\frac{1}{(\phi_0-\eta)^2}\left(k_{10}\eta_0+\eta_{-}+\eta_{+}+\frac{1}{k_1}\eta_n \right) \nonumber \\
	&
	+&\frac{1}{2}\frac{1}{(\phi_0-\eta)^3}\left(k_{10}\eta_0+\eta_{-}+\eta_{+}+\frac{1}{k_1}\eta_n \right)^2,
	\label{contact_hs_ion}
\end{eqnarray} 
where $\phi_0$ is the geometrical porosity of a HS matrix  (\ref{phi0}), $\eta=\sum_{i=-,+,n}\eta_i$, $\eta_{i}=\piup\rho_i\sigma_{i}^3/6$  is the packing fraction of the $i$th species of HS fluid mixture. For the model with a hetero-nuclear cation, $\sigma_-=\sigma_+=\sigma\neq\sigma_n$ and we have:
\begin{equation}
\eta_{-}=\eta_{+}, \quad k_{10}=\sigma_0/\sigma, \quad k_1=\sigma_n/\sigma.
\label{ki}
\end{equation}
Similarly, we obtain the expression for the contact value of the radial distribution function $g_{+n}^{(\text{hs})}(\sigma_{+n})$
\begin{eqnarray}
	g_{+n}^{(\text{hs})}(\sigma_{+n})&=&\frac{1}{\phi_0-\eta}+\frac{3}{2}\frac{1}{(\phi_0-\eta)^2}\left[k_{10}\eta_0+\frac{2}{1+k_1}\left(k_1(\eta_{-}+\eta_{+})+\eta_n\right) \right] \nonumber \\
	&
	+&\frac{1}{2}\frac{1}{(\phi_0-\eta)^3}\left[k_{10}\eta_0+\frac{2}{1+k_1}\left(k_1(\eta_{-}+\eta_{+})+\eta_n\right) \right]^2,
	\label{contact_hs_dimer}
\end{eqnarray}
where the notations are the same as in  (\ref{contact_hs_ion}).

The contact value of the radial distribution function $g_{nn}^{(\text{hs})}(\sigma_{nn}^+)$, has the form:
\begin{eqnarray}
	\label{g22}
	g_{nn}^{(\text{hs})}(\sigma_{nn}^+)&=&\frac{1}{\phi_0-\eta}+\frac{3}{2}\frac{1}{(\phi_0-\eta)^2}\left(k_{10}\eta_0+k_1(\eta_{-}+\eta_{+})+\eta_n \right) \nonumber \\
	&
	+&\frac{1}{2}\frac{1}{(\phi_0-\eta)^3}\left(k_{10}\eta_0+k_1(\eta_{-}+\eta_{+})+\eta_n \right)^2.
	\label{contact_hs_tail}
\end{eqnarray}

For the model of  chain cations  with monomers of the same diameter $k_1=1$. In this case,  from  (\ref{contact_hs_ion})--(\ref{contact_hs_tail}) we have $	g_{+-}^{(\text{hs})}(\sigma)=g_{+n}^{(\text{hs})}(\sigma)=g_{-n}^{(\text{hs})}(\sigma)$.

\subsection{Thermodynamic properties}

We write  the excess internal energy $\Delta E$ of our model at $z=1$ as \cite{Protsykevytch1997,Kalyuzhnyi2001,Kalyuzhnyi2018}
\begin{equation}
	\label{Delta_E}
	\beta\frac{\Delta E}{V}=\frac{\beta e^2}{\varepsilon} \frac{\rho}{\sigma}\left[\left(\sum_{\alpha=0}^{B}X_{+}^{\alpha}-1\right)
	-\left(\sum_{\alpha=0}^BX_{-}^\alpha+1\right)\right].
\end{equation}
It is worth noting that equations (\ref{contact_full}) and (\ref{Delta_E}) include $X_i^{\alpha}$, which, in turn, depend on the parameter~$\Gamma$. 

The free energy of the ionic subsystem $\beta\Delta f$ is presented in the form
of the sum of two terms, namely, the contribution due to bonding or the so-called
mass action law (MAL)  \cite{Bernard96,Jiang02}
and the contribution due to electrostatic interactions:
\begin{equation}
	\beta\Delta f=\beta\Delta f^{(\text{MAL})}+\beta\Delta f^{(\text{el})},
	\label{Delta_F}
\end{equation}
where
\begin{equation}
	\beta\Delta f^{(\text{MAL})}=\rho\left(\ln{x}-{1\over 2}x+{1\over 2}\right)
	-\rho\sum_{i=2}^{m-1}\ln{\left[g_{i,i+1}^{(\text{hs})}(\sigma_{i,i+1})\right]}.
	\label{f-mal}
\end{equation}
Here, $x$ is the fraction of free anions (or cations). $g_{i,i+1}^{(\text{hs})}(\sigma_{i,i+1})$ is the contact values of the radial distribution functions between monomers in the chain cation [see~(\ref{contact_hs_dimer})--(\ref{contact_hs_tail})]. For a model with a cation having the form of a hetero-nuclear dimer, the second term in (\ref{f-mal}) is given by (\ref{contact_hs_dimer}). 

According to  \cite{Jiang02} and in the spirit of Wertheim's multidensity thermodynamic perturbation theory~\cite{Wertheim1984,Wertheim1986} we
assume different approximations for different terms in the expression for 
Helmholtz free energy~(\ref{Delta_F}), i.e., we calculate $\Delta f^{(\text{el})}$ 
using the  parameter $\Gamma$ obtained in the complete dissociation limit
($K^{(0)}_{\text{as}}=0$ and $t=0$).
In this approach, the effects connected with ion screening are taken into account exactly, while the effects of ion association are neglected. As a result, $\beta\Delta f^{(\text{el})}$ is presented as follows:
\begin{equation}
	\label{f-el}
	\beta\Delta f^{(\text{el})}=\frac{\beta\Delta E^{(0)}}{V}+\frac{(\Gamma^{(0)})^3}{3\piup},
\end{equation}
where, $\Gamma^{(0)}=\Gamma\vert_{K_{\text{as}}^{(0)}=0}$ and $\Delta E^{(0)}=\Delta E\vert_{K_{\text{as}}^{(0)}=0}$. Accordingly, the equation for $\Gamma^{(0)}$ is obtained from  (\ref{Gamma-2}) if one put $t=0$.
It should be noted that $\Gamma^{(0)}$ contains a contribution related to the presence of the neutral beads of the cation chain, and cannot be reduced to the screening parameter $\Gamma$ in the MSA theory. 

For $\beta\Delta E^{(0)}$ we have:
\begin{equation}
	\label{Delta_E0}
	\beta\frac{\Delta E^{(0)}}{V}=\frac{\beta e^2}{\varepsilon}\frac{\rho}{\sigma}\left(\bar{X}_{+}^0+\bar{X}_{+}^B-\bar{X}_{-}^0-2\right),
\end{equation}
where $\bar{X}_i^{\alpha}={X}_i^{\alpha}\vert_{\Gamma=\Gamma^{(0)}}$.

The pressure and chemical potentials of the ionic subsystem can be obtained using standard thermodynamic relations, namely:
\begin{equation}
	\label{thermod_ion}
	P=-\frac{\partial F}{\partial V},  \qquad
	\rho(\beta\mu_{-}+\beta\mu_{+})=\beta f+\beta P,
\end{equation}
where $ F=fV$.

\paragraph{Thermodynamic functions in the complete
	association limit.}
Consider the  complete association limit  \cite{Jiang02,KALYUZHNYI1998,Kalyuzhnyi2000} when all anions and cations are
dimerized. In this limit $K_{\text{as}}^{(0)}\to\infty$ and  we get [see (\ref{t})--(\ref{eq-x})]
\begin{equation}
	\label{t_CAL}
	t=\frac{1}{2\rho\sigma}.
\end{equation} 
Taking into account this assumption,  we obtain \cite{Kalyuzhnyi2018}:
\begin{eqnarray}
	\label{f_mal-CAL}
	\beta\Delta f^{(\text{MAL})}\vert_{K_{\text{as}}^{(0)}\to\infty}&=&   -\rho\left\{\ln\rho-1+\ln\left[ g_{-+}^{(\text{hs})}(\sigma_{-+})\right]	+\ln\left[ g_{+n}^{(\text{hs})}(\sigma_{+n})\right]
	\right.
	\nonumber \\
	&+&
	\left.
	\sum_{i=4}^{m-1}\ln{\left[g_{i,i+1}^{(\text{hs})}(\sigma_{i,i+1})\right]}
	+\beta U_{-+}^{(C)}(\sigma_{-+})+G_{00}^{(\infty)}(\sigma_{-+})
	\right\},
\end{eqnarray}
where
\begin{equation}
	\label{G00_CAL}
	G_{00}^{(\infty)}(\sigma_{-+})=G_{00}(\sigma_{-+})\vert_{K_{\text{as}}^{(0)}\to\infty}=\frac{1}{T^*}\frac{\left[1-(\eta^{B(\infty)}\sigma_{-+})^2\right]}{(1+\sigma_{-+}\Gamma^{(\infty)})^2},
\end{equation}
\[
\Gamma^{(\infty)}=\Gamma\vert_{K_{\text{as}}\to\infty}, \qquad
\eta^{B(\infty)}=\Gamma\vert_{K_{\text{as}}\to\infty}.
\]
Here, $\Gamma^{(\infty)}$ is obtained from  (\ref{Gamma-2}) under the condition (\ref{t_CAL}). In the same way, we obtain  $\eta^{B(\infty)}$.

As the result, the pressure of the original matrix-IL system in the complete association limit can be presented as follows: 
\begin{equation}
	\label{P_full_CAL}
	\beta P=\rho+\beta P^{(\text{ref})}+\beta\Delta P^{(\text{MAL})}+\beta\Delta P^{(\text{el})},
\end{equation}
where $\beta P^{(\text{ref})}$ is given by  (\ref{pressureCS1})--(\ref{v-s-q}). Taking into account    (\ref{thermod_ion}) and (\ref{f_mal-CAL}),  we obtain for $\beta \Delta P^{(\text{MAL})}$
\begin{eqnarray}
	\label{P_MAL-CAL}
	\beta\Delta P^{(\text{MAL})}&=&-\rho^2\left[\frac{\partial \ln[ g_{-+}^{(\text{hs})}(\sigma_{-+})]}{\partial \rho} + \frac{\partial \ln[ g_{+n}^{(\text{hs})}(\sigma_{+n})]}{\partial \rho}\right. \nonumber \\
	&+&
	\left.
	\sum_{i=4}^{m-1}
	{\partial\ln{\left[g_{i,i+1}^{(\text{hs})}(\sigma_{i,i+1})\right]}\over\partial\rho}
	+\frac{\partial G_{00}^{(\infty)}(\sigma_{-+})}{\partial \rho}  \right],
\end{eqnarray}
where $g_{-+}^{(\text{hs})}(\sigma_{-+})$, $g_{+n}^{(\text{hs})}(\sigma_{+n})$, $g_{i,i+1}^{(\text{hs})}(\sigma_{i,i+1})$,  and $G_{00}^{(\infty)}(\sigma_{-+})$ are given in (\ref{contact_hs_ion})--(\ref{contact_hs_tail}), and (\ref{G00_CAL}), respectively.

Electrostatic contribution $\beta\Delta P^{(\text{el})}$ is
\begin{equation}
	\label{P0_el}
	\frac{\beta\Delta P^{(\text{el})}}{\rho}=-\frac{(\Gamma^{(0)})^3}{3\piup\rho}-\frac{2\beta e^2}{\piup\varepsilon\rho}(\eta_0^B)^2.
\end{equation} 

Similar to (\ref{P_full_CAL}), we represent the chemical potential $\beta\mu=(\beta\mu_{-} +\beta\mu_{+})$ in the form:
\begin{equation}
	\label{mu_full_CAL}
	\beta\mu=\ln\rho+\beta\mu^{(\text{ref})}+\beta\Delta\mu^{(\text{MAL})}+\beta\Delta\mu^{(\text{el})},
\end{equation}
where $\beta\mu^{(\text{ref})}=(\beta\mu_{-}^{\text{ref}} +\beta\mu_{+}^{\text{ref}})$  can be obtained from (\ref{mu_ref})--(\ref{muCS}).  
For $\beta\Delta\mu^{(\text{MAL})}$, using (\ref{thermod_ion}), we have
\begin{equation}
	\label{mu_MAL-CAL}
	\beta\Delta\mu^{(\text{MAL})}=\frac{\beta\Delta f^{(\text{MAL})}}{\rho}+\frac{\beta\Delta P^{(\text{MAL})}}{\rho},
\end{equation}
where $\beta\Delta f^{(\text{MAL})}$ and $\beta\Delta P^{(\text{MAL})}$ are given in (\ref{f_mal-CAL}) and (\ref{P_MAL-CAL}), respectively. The contribution  $\beta\Delta\mu^{(\text{el})}$ has the form:
\begin{equation}
	\label{mu_el-CAL}
	\beta\Delta\mu^{(\text{el})}=\beta\frac{\Delta E^{(0)}}{V\rho}-\frac{2\beta e^2}{\piup\varepsilon\rho}(\eta_0^B)^2.
\end{equation}


\section{Results and discussion} \label{sec3}
Using the theory proposed in the previous section, we have studied 
the vapour-liquid phase behaviour of two models of ILs, model~A and model~B,  confined in a disordered matrix  in the approximation of complete ion association [see equations~(\ref{P_full_CAL})--(\ref{mu_el-CAL})]. 

The phase diagrams were obtained at subcritical temperatures using the conditions of two-phase equilibrium
\begin{eqnarray*}
	\mu(\rho^{(\text{v})},T)&=&\mu(\rho^{(\text{l})},T), 
	\\ 
	P(\rho^{(\text{v})},T)&=&P(\rho^{(\text{l})},T), 
\end{eqnarray*} 
where the subscripts ``v'' and ``l'' refer to the vapour and the liquid phases, respectively.

We introduce dimensionless units for the temperature, pressure,  cation (anion) number density, and volume fraction
\begin{eqnarray*}
	T^*=\frac{k_{\text{B}}T\varepsilon\sigma}{\rm{e}^2}, \quad
	P^*=\frac{P\varepsilon\sigma^4}{\rm{e}^4}, \quad
	\rho^*=\rho\sigma^3, \quad
	\eta=\frac{\piup}{6}\rho\sum_{\alpha=-,+,n}\sigma_{\alpha}^3,
\end{eqnarray*}
where  $\sigma=\sigma_{-}=\sigma_{+}$ is the diameter of the charged monomer  which is the same for anion and cation,
and $\sigma_n$ is the diameter of the uncharged monomer of the cation. The ionic model is characterised  by  the parameter $k_1=\sigma_n/\sigma$ [equation~(\ref{ki})].

\begin{figure}[t]
	\begin{center}
		\includegraphics[width=0.45\textwidth]{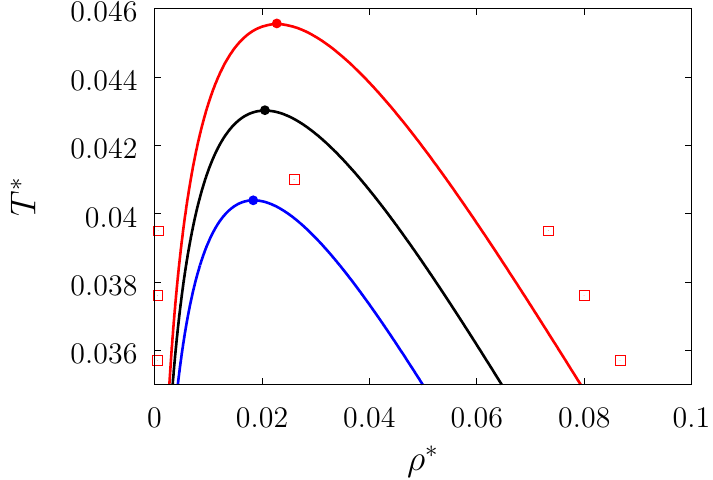}
		\includegraphics[width=0.45\textwidth]{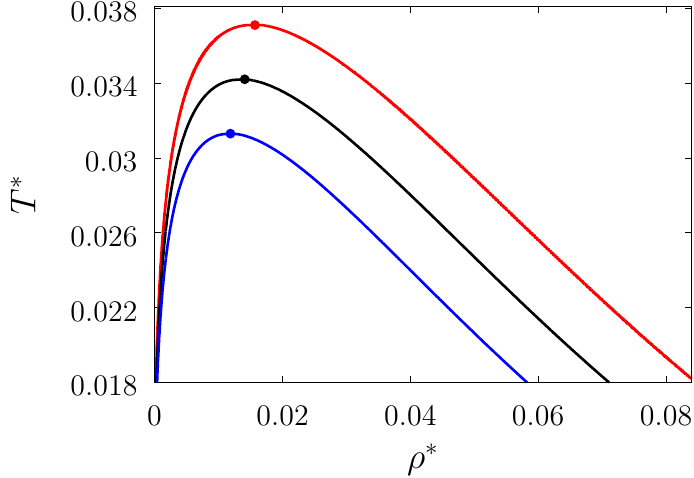}\\
		\includegraphics[width=0.45\textwidth]{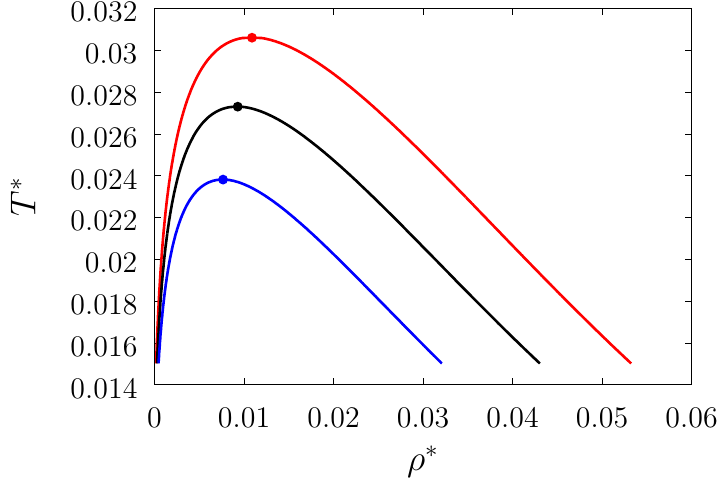}
		\includegraphics[width=0.45\textwidth]{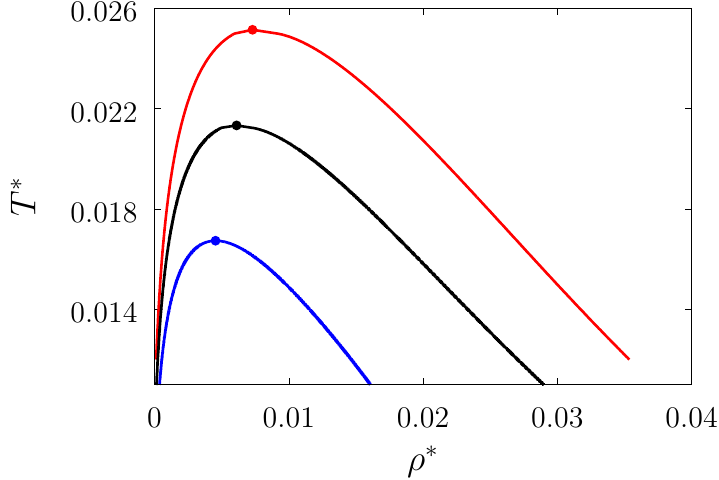}
		\caption{(Colour online) Vapour-liquid phase diagrams of model~A with   $k_1=1$ (upper left-hand panel),  $k_1=1.5$ (upper right-hand panel),  $k_1=2$ (lower left-hand panel),  $k_1 =2.5$ (lower right-hand panel), confined in a porous matrix with different porosity. From the top to the bottom: $\phi_0=1.0$ (red lines and squares), $\phi_0=0.95$ (black lines), $\phi_0=0.9$ (blue lines). The diameter of matrix particles $\sigma_0=1.5\sigma$. Here, the lines represent theoretical results, and the symbols are the results of computer simulations \cite{Kalyuzhnyi2018}. Notations:  $T^*=k_{\text{B}}T\varepsilon\sigma/e^2$, $\rho^*=\rho\sigma^3$, $k_1=\sigma_n/\sigma$,   and $\phi_0=1-\eta_0$, where $\eta_0=\piup\rho_0\sigma_0^3/6$. }
		\label{fig_1}
	\end{center}
\end{figure}
\subsection{Model~A}
First, we present the results for model~A (see figure~\ref{models}). 
Calculations are done for three values of the matrix porosity  $\phi_0$, i.e., $\phi_0=1$ (bulk case), $\phi_0=0.95$ and $\phi_0=0.9$. We fix the diameter of the matrix particles $\sigma_0$ and study the effect of the porous medium on the phase behavior of model~A with different $k_1$.  
In figure~\ref{fig_1},  we present the vapour-liquid phase diagrams of model~A   confined in a disordered matrix with the diameter of matrix particles $\sigma_0=1.5$. Comparing the results obtained for different  $k_1$ in the absence of a porous matrix, the case $\phi_0=1$ in figure~\ref{fig_1} (red curves), we see that an increase of $k_1$  leads to a shift of the phase diagrams to lower temperatures and to lower number densities which agrees with the results obtained in \cite{Kalyuzhnyi2018}. For  the fixed porosity $\phi_0<1$ (black and blue curves), an increase of $k_1$  also has a similar effect: the coexistence curves shift to lower temperatures and densities  when  $k_1$ increases, and it can be concluded that the presence of the porous medium strengthens the effects previously observed for the bulk case.
On the other hand, for the fixed  $k_1$, a decrease of matrix porosity $\phi_0$ also leads to a shift of the phase diagram to lower  temperatures $T^*$ and to lower densities $\rho^*$, and  the phase coexistence region becomes narrower. For comparison, figure~\ref{fig_1} (upper left-hand panel) shows the results of computer simulations for $k_1=1$ in the absence of a porous medium. It should be noted that the agreement of our theoretical results for the IL model in the absence of a porous matrix with the results of computer simulations is within the same order of accuracy as the theoretical predictions made earlier for the RPM model (see~\cite{Kalyuzhnyi2018}).
\begin{figure}
	\begin{center}
		\includegraphics[width=0.46\textwidth]{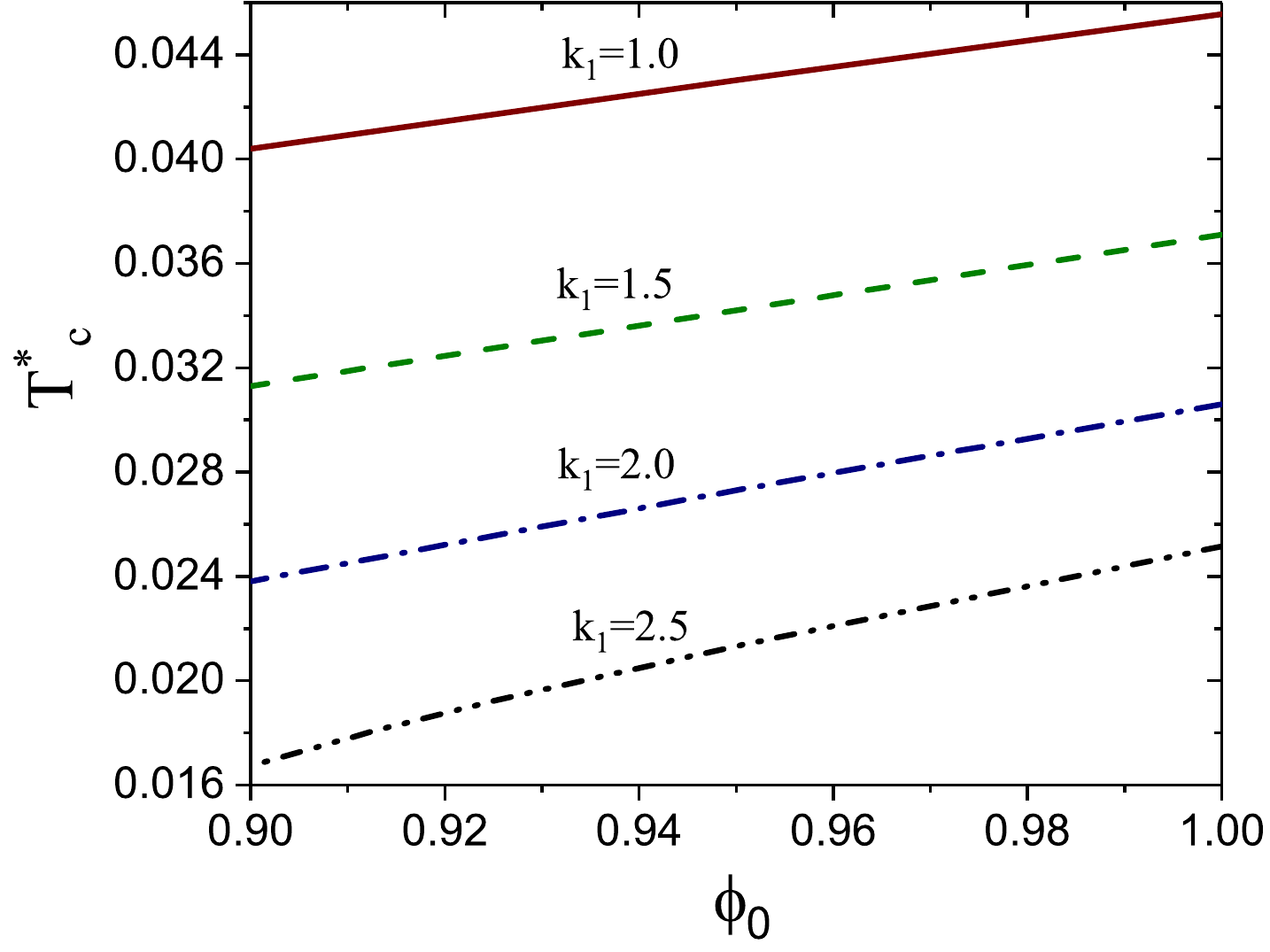}
		\caption{(Colour online) Vapour-liquid critical temperature $T_{c}^*$ of  model~A confined in  the porous medium as a function of the porosity $\phi_0$. From the top to the bottom: $k_1=1$, $k_1=1.5$, $k_1=2$, $k_1 =2.5$.  Notations: $T_c^*=k_{\text{B}}T_c\varepsilon\sigma/e^2$, $k_1=\sigma_n/\sigma$ and $\phi_0=1-\eta_0$, where $\eta_0=\piup\rho_0\sigma_0^3/6$. }
		\label{fig_2}
	\end{center}
\end{figure}
\begin{figure}[!t]
	\begin{center}
		\includegraphics[width=0.45\textwidth]{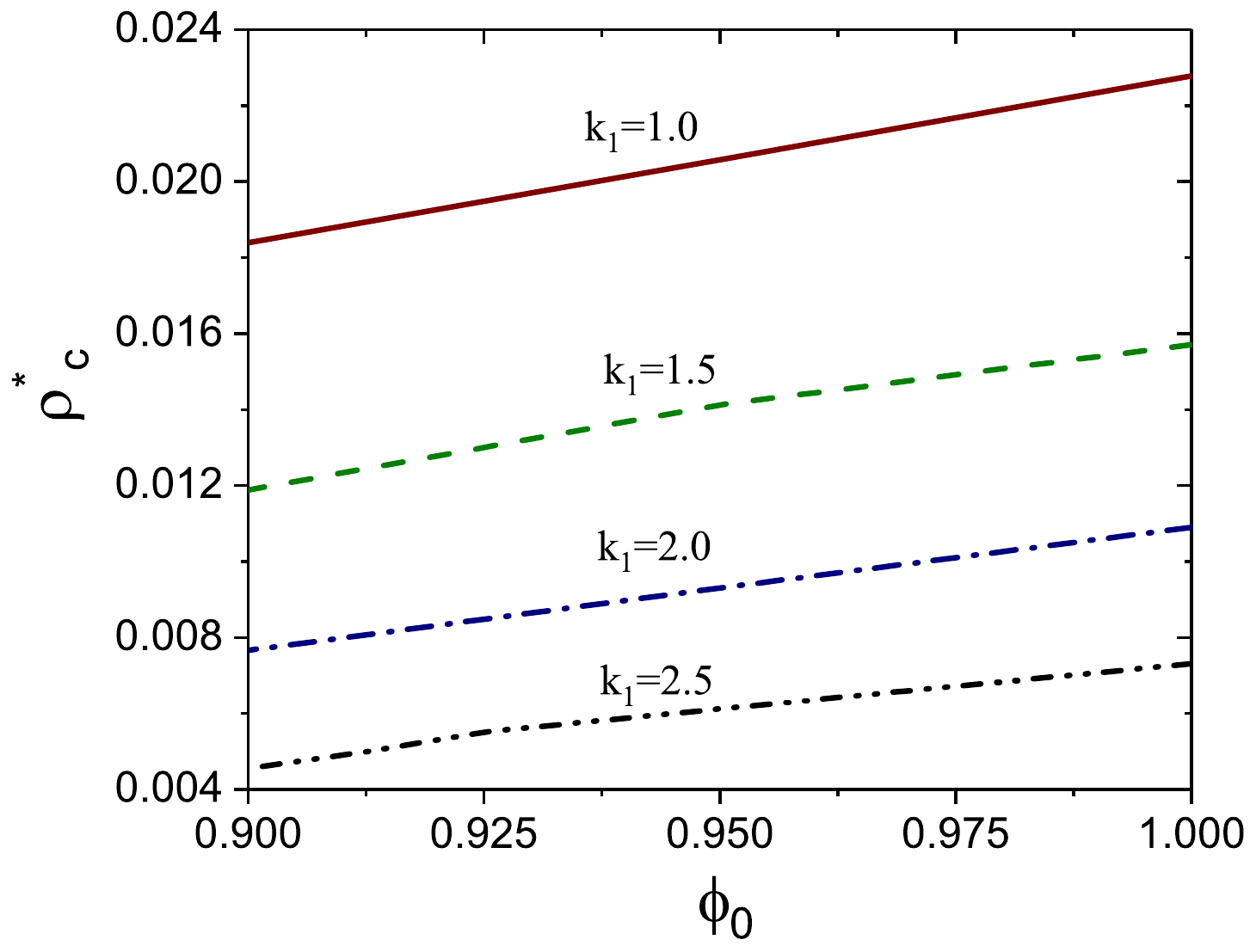}
		\includegraphics[width=0.45\textwidth]{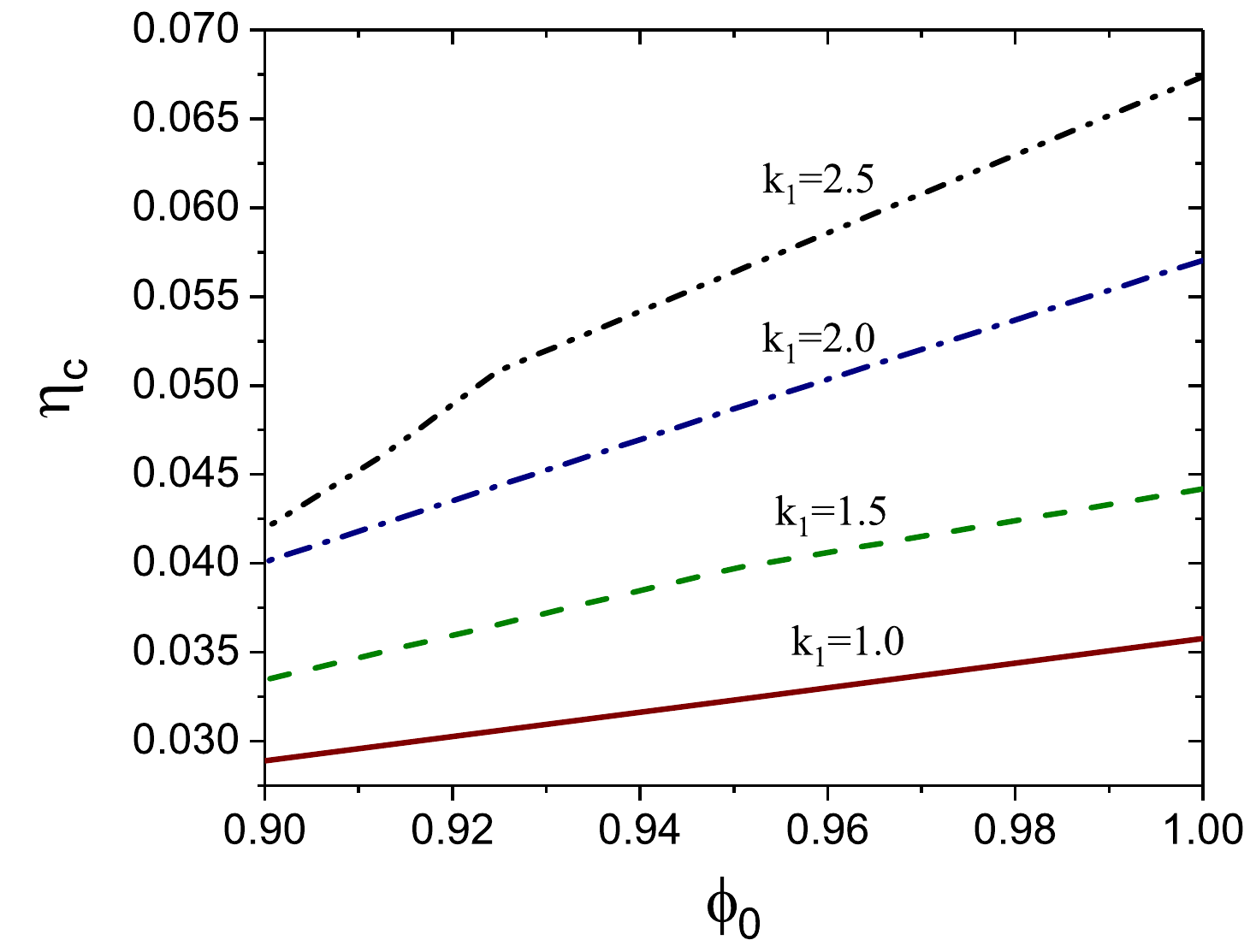}
		\caption{(Colour online) Critical density $\rho_c^*$ (left-hand panel) and critical volume fraction $\eta_c$ (right-hand panel) of model~A confined in  the porous medium as a function of the porosity $\phi_0$ at different values of $k_1$ (as indicated on  the plots). Notations: $\rho_c^*=\rho_c\sigma^3$, $\eta_c=\piup\rho_c\sum_{\alpha=-,+,n}\sigma_{\alpha}^3/6$, $k_1 =\sigma_n/\sigma$, and $\phi_0=1-\eta_0$, where $\eta_0=\piup\rho_0\sigma_0^3/6$. }
		\label{fig_3}
	\end{center}
\end{figure}
\begin{figure}[!t]
	\begin{center}
		\includegraphics[width=0.5\textwidth]{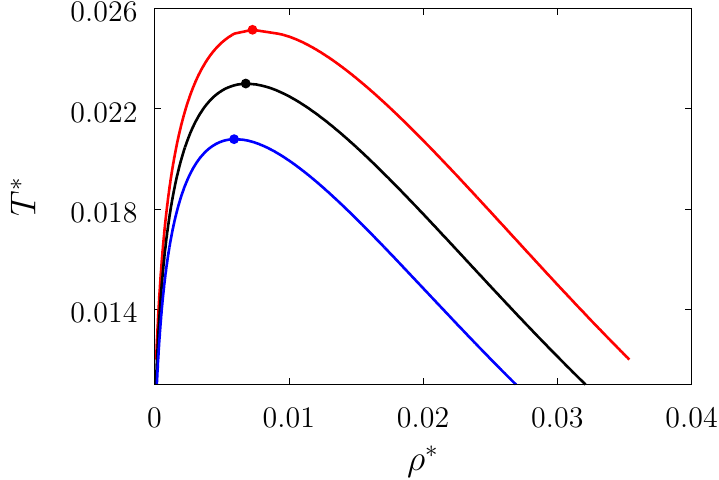}
		\caption{(Colour online) Vapour-liquid phase diagrams of  model~A with $k_1=2.5$, confined in a porous matrix with a matrix porosity $\phi_0=1.0$ (red curves ), $\phi_0=0.95$ (black curves), $\phi_0=0.9$ (blue curves).  The diameter of matrix particles  $\sigma_0=2.5\sigma$. Notations: $T^*=k_{\text{B}}T\varepsilon\sigma/e^2$, $\rho^*=\rho\sigma^3$, $k_1=\sigma_n/\sigma$ and $\phi_0= 1-\eta_0$, where $\eta_0=\piup\rho_0\sigma_0^3/6$.}
		\label{fig_5}
	\end{center}
\end{figure}

The dependence of the critical  temperature $T_c^*$ on the matrix  porosity  $\phi_0$ for  $k_1=1$, $1.5$, $2$ and $2.5$ is shown in  figure~\ref{fig_2}. It is seen that the critical temperature $T_c^*$ decreases almost linearly with a decreasing porosity for all  considered values of $k_1$, although a small deviation from linearity can be observed for $k_1=2.5$. Moreover, the critical temperature $T_c^*$ decreases with an increase of $k_1$.

In figure~\ref{fig_3}, we show the dependence of the critical density $\rho_c^*$ and the critical volume fraction  $\eta_c$ {($\eta_c=\piup\rho_c\sum_{\alpha=-,+,n}\sigma_{\alpha}^3/6$)} on the matrix porosity  $\phi_0$. 
Again, the critical density $\rho_c^*$ decreases almost linearly with a decreasing matrix porosity for all values of $k_1$ except for $k_1=2.5$, for which some deviation from a linear behavior is observed: a decrease of $\rho_c^*$ becomes steeper for $\phi_0\leqslant 0.925$. Similar to  $T_c^*$, the critical density $\rho_c^*$ decreases when $k_1$ increases. {It should be noted that a similar behavior of the critical density with a decreasing porosity  was obtained for a size-asymmetric primitive model of  spherical ions  if the difference between the ion diameters increases  \cite{patsahan2018vapor}.}  Regarding the critical  volume fraction $\eta_c$, its behavior with a decrease of matrix porosity is similar to the behavior of  $\rho_c^*$, except that the deviation from the linear dependence of $\eta_c$ is more pronounced for $ \phi_0\leqslant 0.925$. On the other hand, the dependence of $\eta_c$ on $k_1$ is opposite to that obtained for $\rho_c^*$, i.e., when $k_1$ increases $\eta_c$ also increases.

In figure~\ref{fig_5}, we show the vapour-liquid phase diagrams obtained for   model~A with $k_1=2.5$ confined in a porous matrix with the matrix particles of the diameter 
$\sigma_0=2.5\sigma$. Comparing these phase diagrams with the corresponding phase diagrams in figure~\ref{fig_1}~(lower right-hand panel) for $\sigma_0=1.5\sigma$, one can see that an increase of the diameter of the matrix particles for the fixed matrix porosity  $\phi_0$ leads to a shift of the phase diagrams to higher temperatures $T^*$ and to higher densities $\rho^*$, as well as to the broadening of the range of phase coexistence.
\begin{figure}[!t]
	\begin{center}
		\includegraphics[width=0.45\textwidth]{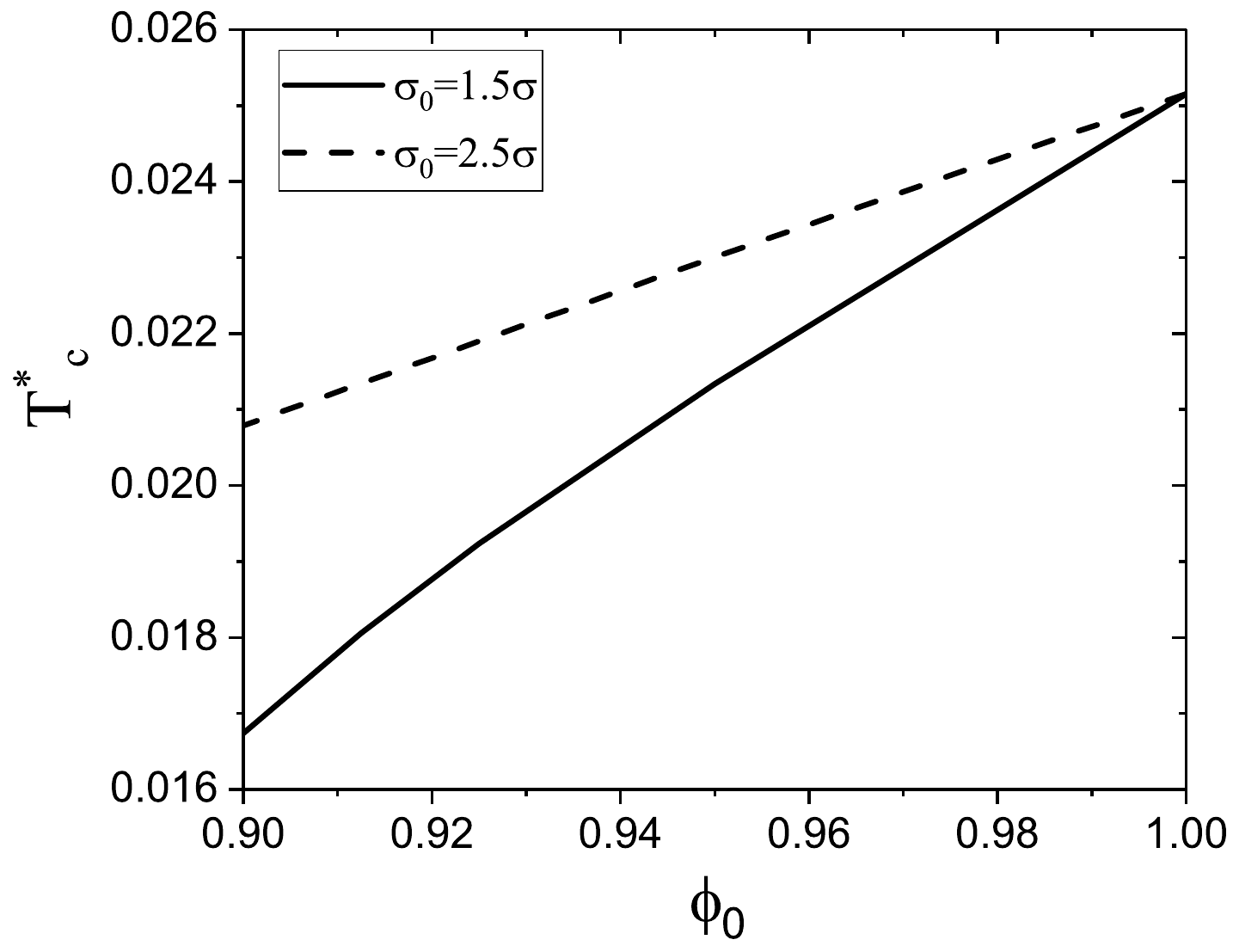}
		\includegraphics[width=0.45\textwidth]{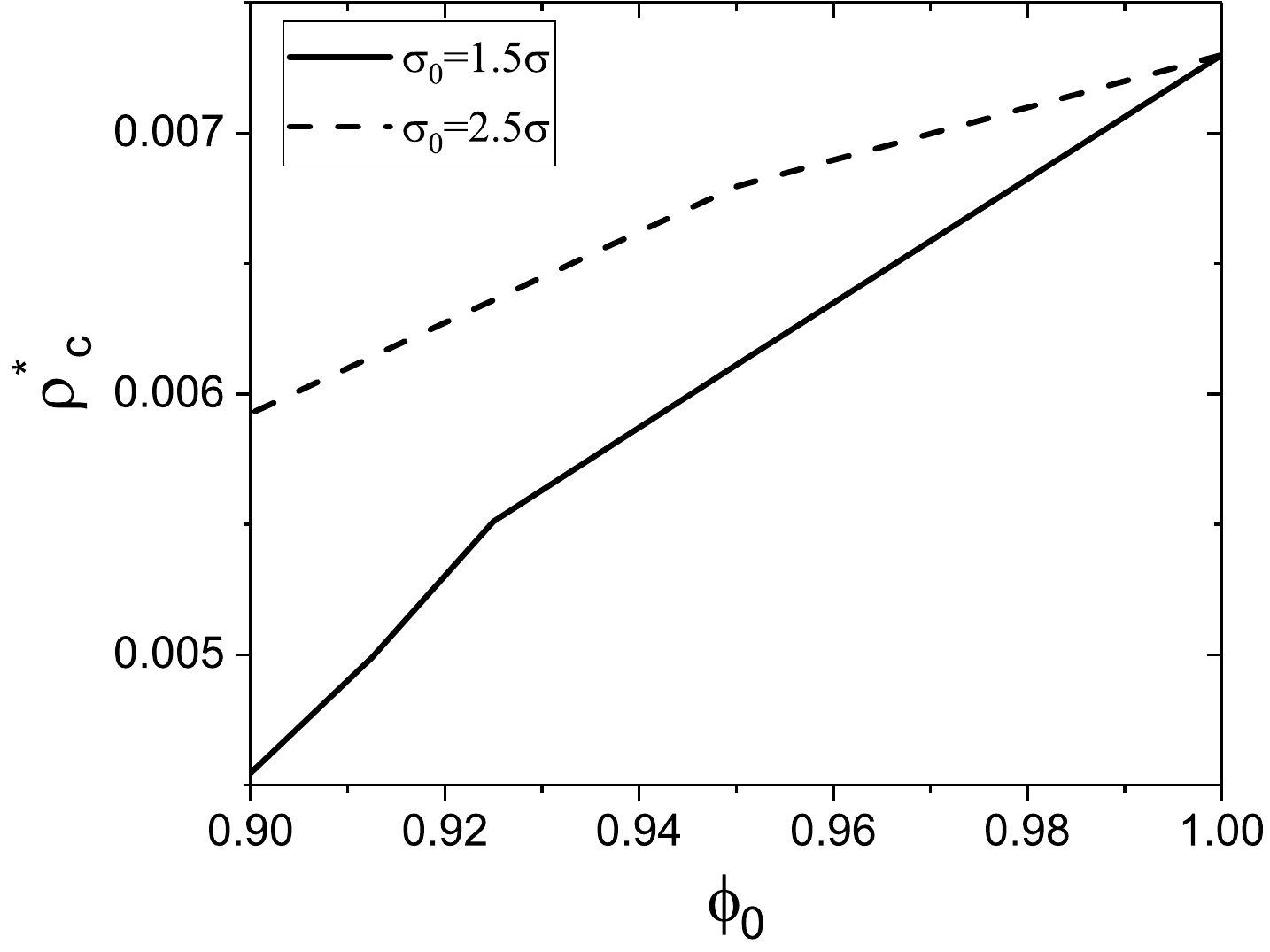}
		\caption{Critical temperature $T_{c}^*$ (left-hand panel)  and the critical density  $\rho_{c}^*$  (right-hand panel) of  model~A with $k_1=2.5$ confined in a porous matrix as a function of matrix porosity $\phi_0$ for two sizes of matrix particles: $\sigma_0=1.5\sigma$ (solid curve) and $\sigma_0=2.5\sigma$ (dashed curve). Notations: $T_c^*=k_{\text{B}}T_c\varepsilon\sigma/e^2$, $k_1=\sigma_n/\sigma$ and $\phi_0=1-\eta_0$, where $\eta_0=\piup\rho_0\sigma_0^3/6$. }
		\label{fig_6}
	\end{center}
\end{figure}

The dependence of the critical temperature $T_{c}^*$  and the critical number density  $\rho_c^*$ for  model~A with  $k_1=2.5$ on the matrix porosity $\phi_0$ for two   diameters of matrix particles  $\sigma_0=1.5\sigma$ (solid curve) and $\sigma_0=2.5\sigma$ (dashed curve) is shown in figure~\ref{fig_6}. It is seen that
an increase of the diameter of  matrix obstacles  does not generally change the behavior of $T_{c}^*$  and $\rho_c^*$: both $T_{c}^*$ and $\rho_c^*$ decrease with a decrease of the matrix porosity, as it was observed in the case of $\sigma_0=1.5\sigma$.
However, the decrease of  both critical parameters is much slower for $\sigma_0=2.5\sigma$ and the difference between $T_{c}^*$  ($\rho_c^*$) obtained for $\sigma_0=1.5\sigma$ 
and $\sigma_0=2.5\sigma$ increases with a decreasing porosity. 
Such a behaviour of the critical parameters with the change of the size of matrix particles  is consistent with the results obtained earlier for the RPM model~\cite{holovko2016vapour,holovko2017effects} {as well as with the results for a size-asymmetric primitive models \cite{patsahan2018vapor,HolPatPat18charge}.}

\begin{figure}[!t]
	\begin{center}
		\includegraphics[width=0.46\textwidth]{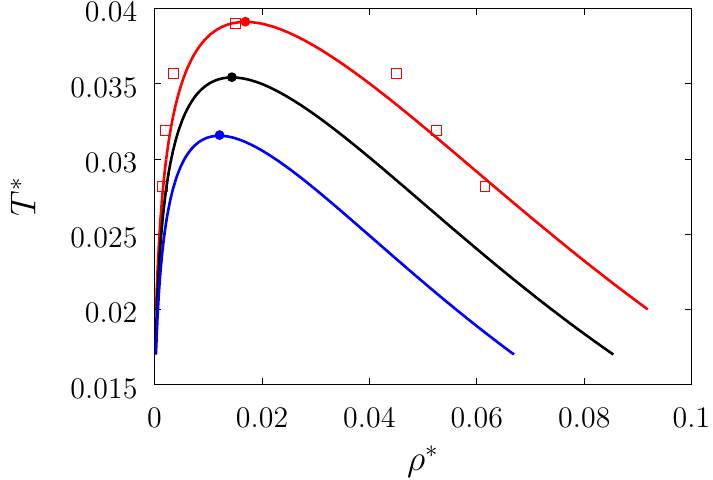} \\
		\includegraphics[width=0.46\textwidth]{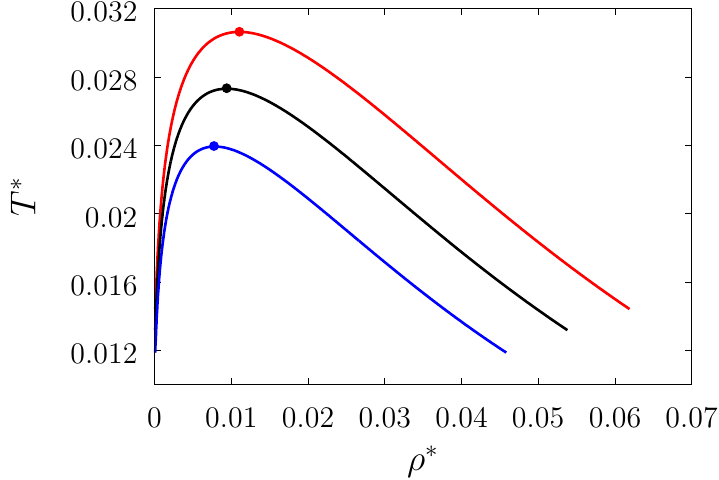}
		\includegraphics[width=0.46\textwidth]{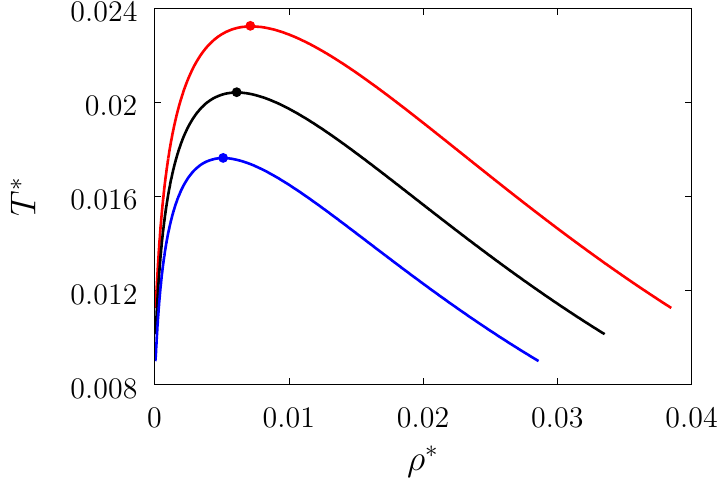}
		\caption{(Colour online) Vapour-liquid phase diagrams of  model~B with cation chain length:  $m_+=3$ (upper panel), $m_+=5$ (lower left-hand panel), $m_+=8$ (lower right-hand panel), confined in a porous matrix with porosity $\phi_0=1.0$ (red lines and squares), $\phi_0=0.925$ (black lines), $\phi_0=0.85$ (blue lines). Here, the lines represent theoretical results, and the symbols are the results of computer simulation \cite{Kalyuzhnyi2018}. Notations: $T^*=k_{\text{B}}T\varepsilon\sigma/e^2$ and $\rho^*=\rho\sigma^3$. }
		\label{fe_n}
	\end{center}
\end{figure}

\subsection{{Model~B}}
\begin{figure}[!t]
	\begin{center}
		\includegraphics[width=0.46\textwidth]{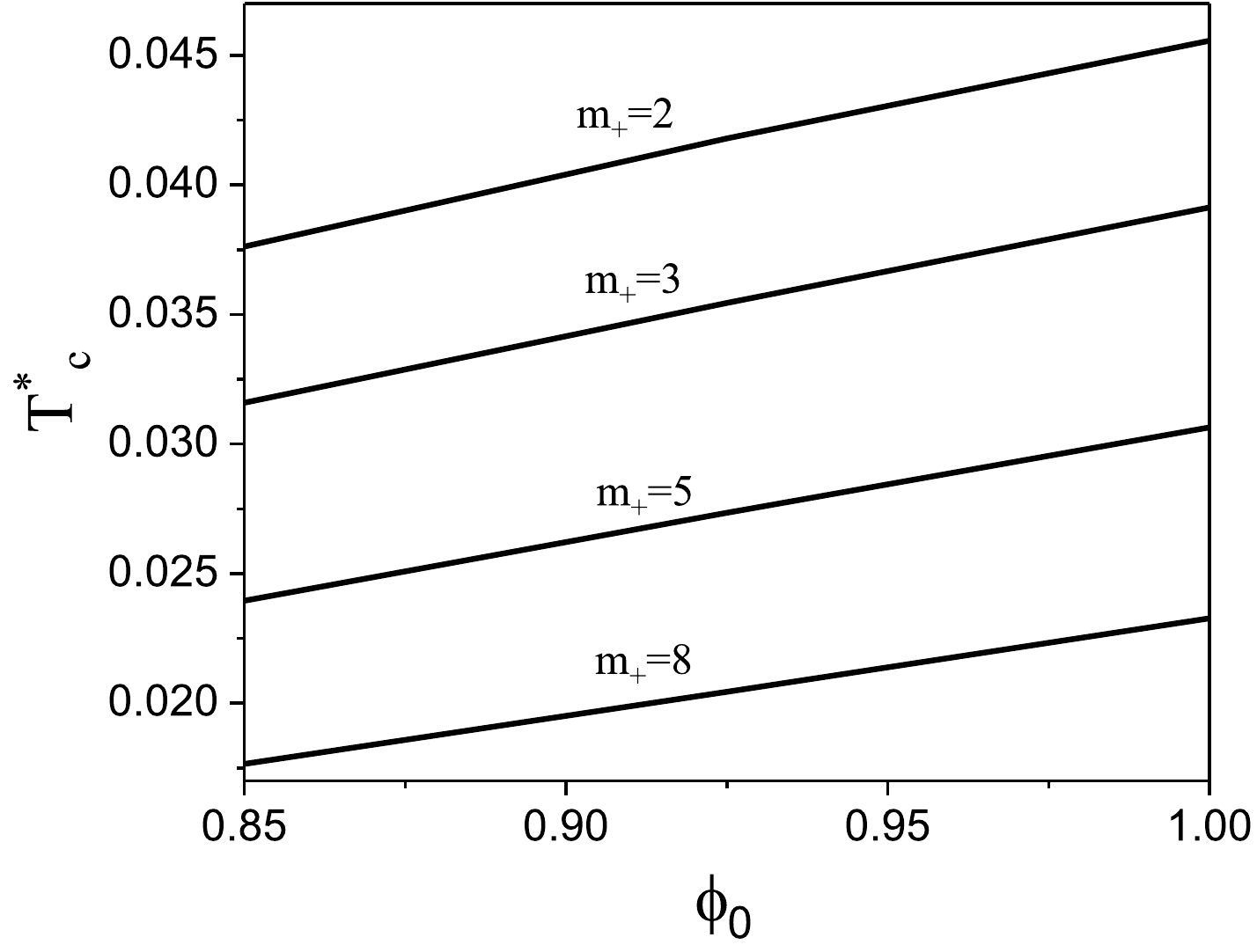}
		\includegraphics[width=0.46\textwidth]{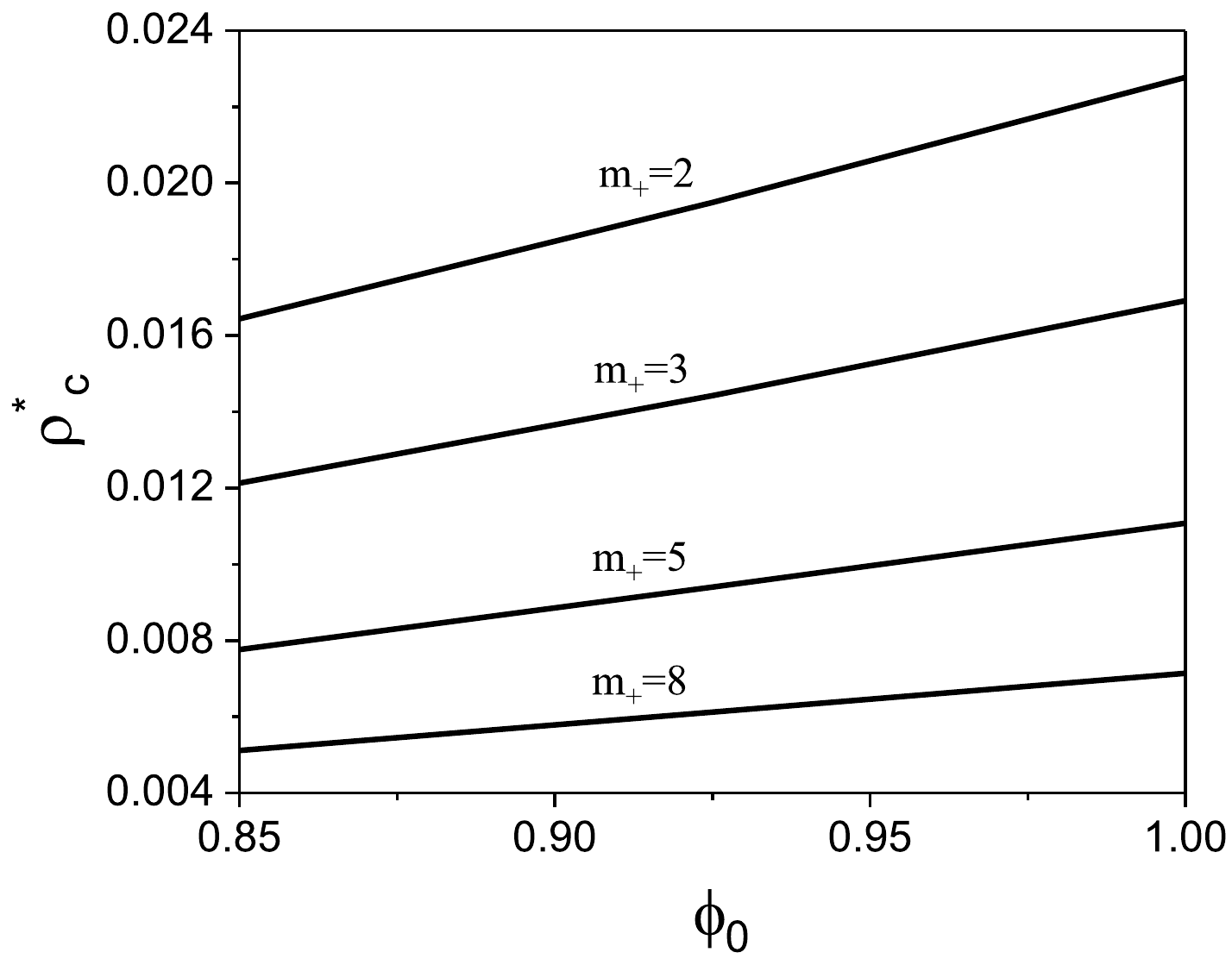}
		\caption{Dependence of the critical temperature (left-hand panel) and the critical density (right-hand panel) on the matrix porosity $\phi_0$ for  model~B with different lengths of the cation chains $m_{+}=2,3,5,8$ confined in a porous medium. Notations are the same as in figure~\ref{fe_n}.}
		\label{crit_par-phi0}
	\end{center}
\end{figure}

 Now we focus on model~B. In this model, the cations are presented as mononuclear chains  with $m_{+}$  beads (figure~\ref{models}). The vapour-liquid phase diagrams are calculated for four values of the length of the cation chain $m_{+}=m-1=3,5,8$ and for three values of the matrix porosity  $\phi_0=1.0, 0.925, 0.85$. In our calculations, we fix the diameter of matrix particles $\sigma_0=1.5$. The phase diagrams are shown in figure~\ref{fe_n}. In the upper panel of these figures, in addition to the theoretical results, we also show
the  simulation results for $m_{+}= 3$ and $\phi_0=1$ (bulk case)  \cite{Kalyuzhnyi2018}.
In this case,  our theoretical results for the vapour-liquid phase diagram are in a good agreement
with the results obtained  using the Monte Carlo simulations. Moreover, the agreement for the cation with $m_{+}= 3$ is much better than in the case of mononuclear dimer corresponding to $m_{+}= 2$  (see figure~\ref{fig_1}, upper left-hand panel).  We can expect that  in the presence of a porous medium and longer cation chains, our results are satisfactory. As for  model~A, the presence of a porous medium  shifts the phase diagrams  of model~B to lower temperatures and to lower number densities when the matrix porosity decreases (figure~\ref{fe_n}). Both the critical temperature and the critical number density decrease almost linearly when the matrix porosity decreases (figure~\ref{crit_par-phi0}). An increase of  the cation chain length   in the absence of a porous medium  ($\phi_0=1.0$) has a similar effect. The presence of the porous matrix significantly enhances this effect.

It should be noted that there is a small difference between the present results for the bulk case and the results obtained in \cite{Kalyuzhnyi2018}. Comparing the results, one can see that the critical temperatures for $\phi_0=1$ in figure~\ref{fig_1} (upper left-hand panel) and figure~\ref{fe_n} (upper panel)  are about $3$\%  higher than the corresponding critical temperatures obtained for the same models in \cite{Kalyuzhnyi2018}.  The difference appears due to different approximations used to calculate the contact values, i.e., in this work the expressions for the contact values of the radial distribution functions [equations~(\ref{contact_hs_ion})--(\ref{contact_hs_tail})] include the term corresponding to the Boublik-Mansoori-Carnahan-Starling-Leland approximation  while in \cite{Kalyuzhnyi2018} the Percus-Yevick contact values are used.  Here, contrary to \cite{Kalyuzhnyi2018}, 
we use the same approximation for the thermodynamic functions of the RS and for the contact values.

\section{Conclusions}\label{sec4}

We have studied the  vapour-liquid phase behaviour of ILs with complex molecular structures of ions (cations) confined in a disordered
matrix formed by neutral HSs. The IL is modelled as an electroneutral mixture  of 
HS anions and  flexible linear chain cations, formed by the tangentially bonded HSs with the charge located on one of the terminal beads \cite{Kalyuzhnyi2018}.
To this end, we have developed  the theoretical approach which combines  a generalization of the scaled particle theory (SPT), Wertheim's thermodynamic perturbation theory, and the associative mean-spherical approximation (AMSA).
Using this theory, we obtained analytical expressions for the Helmholtz free energy, pressure, and chemical potentials of ions in the complete association limit. The thermodynamic functions contain contributions from the RS,  represented by a multicomponent HS fluid of different sizes, confined in the HS matrix, and the ionic subsystem, represented by spherical anions and chain cations. It should be noted that the contribution of the ionic subsystem to the thermodynamics of the full IL-matrix system depends on the contact values of the radial distribution functions of the RS, particularly, on the contact values of the radial distribution functions between the HSs belonging to the anion and the cation and on the contact values of the radial distribution function between the HSs belonging only to the cation. 

We have calculated the vapour-liquid phase diagrams of two versions of the above described IL model, i.e., models with cations represented by dimers with charged and neutral beads of different size and by chains with  neutral beads of the same size, respectively.
Our results show that a decrease in the geometric porosity leads to a shift of the phase diagrams to lower temperatures and to lower values of the number density of anions (cations), as well as to lower values of volume fractions of IL for all considered models of the cation, while the region of the coexistence of the two phases narrows.  This agrees with the results obtained for both simple (nonionic) liquids and model ILs with spherical ions. On the other hand, the results for IL models with molecular cations represented by dimers with the neutral bead of different size and by chains with the neutral beads of the same size
in the bulk  showed \cite{Kalyuzhnyi2018} that an increase of the diameter of the neutral bead or chain length also leads to a shift of the vapour-liquid phase diagram to lower temperatures and densities. Therefore,  the presence of a disordered porous medium enhances these effects  on the vapour-liquid phase diagram of ILs. Accordingly, the critical temperature and critical number density  decrease when the matrix porosity decreases and the diameter/length of the neutral part of the cation increases.  An increase in the size of the matrix particles at a fixed matrix porosity leads to the opposite effect, i.e., the critical temperature $T_c^*$ and  the critical number density $\rho_c^*$  increase with an increase of the diameter of the matrix particles and, accordingly, the phase diagram shifts to higher values of temperature and density.

It should be noted that in this paper the thermodynamic functions of the reference system and the contact values of the radial distribution functions are treated within the framework of the same approximation. This leads, in general, to a bit better agreement between the theoretical and simulation results 
in the bulk case compared to \cite{Kalyuzhnyi2018}. We hope that this is also valid in the presence of a disordered matrix.

\section*{Acknowledgements}
We thank the  Ministry of Education and Science of Ukraine for its financial support (Agreement No~PH/16-2023).
YVK acknowledge financial support through the MSCA4Ukraine project, which is funded by	the European Union.


\bibliographystyle{cmpj}
\bibliography{prepr_refs}

\ukrainianpart

\title{Фазова поведінка газ-рідина примітивних моделей іонних рідин у невпорядкованому пористому середовищі}
\author{Т. Гвоздь\refaddr{label1}, Т. Пацаган\refaddr{label1,label2}, Ю. Калюжний\refaddr{label1,label3}, О, Пацаган\refaddr{label1}, М. Головко\refaddr{label1}}
\addresses{
\addr{label1} Iнститут фiзики конденсованих систем Нацiональної академiї наук України
79011, м. Львiв, вул. Свєнцiцького, 1, Україна
\addr{label2} Iнститут прикладної математики та фундаментальних наук, Нацiональний унiверситет “Львiвська
полiтехнiка”, 79013, Львiв, вул. С. Бандери, 12, Україна,
\addr{label3} Факультет хiмiї i хiмiчної технологiї, Унiверситет Любляни, вул. Вечна, 113, 1000 Любляна, Словенiя
}
%
%
%

\makeukrtitle

\begin{abstract}
\tolerance=3000%
Ми розробляємо теорію для опису іонних рідин у пористому середовищі, сформованому матрицею
нерухомих хаотично розміщених незаряджених частинок. Іонна рідина моделюється як
електронейтральна суміш аніонів сферичної форми та ланцюговoподібних катіонів, представлених
тангенціально зв'язаними твердими сферами із зарядом, розташованим на одній із кінцевих сфер. Теорія
поєднує в собі узагальнення теорії масштабної частинки, термодинамічної теорії збурень Вертгайма та
асоціативного середньо-сферичного наближення і дозволяє отримати аналітичні вирази для тиску та
хiмiчних потенцiалiв системи ``матриця–іонна рідина''. Використовуючи цю теорію, ми розрахували
фазові діаграми ``газ-рiдина'' для двох версій моделі іонної рідини, а саме, коли катіон моделюється як
димер і як ланцюг, в наближенні повної асоціації. Досліджено вплив невпорядкованої матриці та
несферичної форми катіонів на фазові діаграмах ``газ-рiдина''.
\keywords іонні рідини, невпорядковане пористе середовище, ланцюговоподібні катіони, фазові діаграми ``газ-рiдина''

\end{abstract}

\end{document}